\documentclass[a4paper,11pt,aps,preprintnumbers,amsmath,amssymb,superscriptaddress,nofootinbib]{article}
\pdfoutput=1
\usepackage{jheppub}

\usepackage{graphicx}
\usepackage[normalem]{ulem}
\usepackage{hyperref}
\usepackage{dcolumn}
\usepackage{xcolor}

\usepackage{multirow}

\def\beq{\begin{equation}}
\def\eeq{\end{equation}}
\def\bea{\begin{eqnarray}}
\def\eea{\end{eqnarray}}

\newcommand{\mpt}{{\;/\!\!\!\! \vec{P}_T}}

\def\figureautorefname~#1\null{Fig.\,#1\null}
\def\tableautorefname~#1\null{Tab.\,#1\null}

\def\equationautorefname~#1\null{Eq.\,(#1)\null}
\allowdisplaybreaks[4]

\makeatletter

\title{Direct Higgs-top CP-phase measurement with $t\bar{t}h$ at the 14~TeV LHC and 100~TeV FCC}

\author[a]{Dorival Gon\c{c}alves,}
\author[b,c,d]{Jeong Han Kim\footnotemark[1],}
\author[e]{Kyoungchul Kong,}
\author[,a]{Yongcheng Wu\footnotemark[1]}
\footnotetext[1]{Corresponding authors}
\affiliation[a]{Department of Physics, Oklahoma State University, Stillwater, OK, 74078, USA}
\affiliation[b]{Department of Physics, Chungbuk National University, Cheongju, Chungbuk 28644, Korea}
\affiliation[c]{Center for Theoretical Physics of the Universe,
Institute for Basic Science, Daejeon 34126, Korea}
\affiliation[d]{Korea Institute for Advanced Study (KIAS), School of Physics, Seoul 02455, Korea}
\affiliation[e]{Department of Physics and Astronomy, University of Kansas, Lawrence, KS 66045, USA}

\emailAdd{dorival@okstate.edu}
\emailAdd{jeonghan.kim@cbu.ac.kr}
\emailAdd{kckong@ku.edu}
\emailAdd{ywu@okstate.edu}

\abstract{The study of the Higgs boson's properties is a cornerstone of the LHC and future collider programs. In this paper, we examine the  potential to directly probe the Higgs-top interaction strength and CP-structure in the $t\bar t h$ channel with the Higgs boson decaying to bottom-quark pairs and top-quarks in the di-leptonic mode. We adopt the BDRS algorithm to tag the boosted Higgs and exploit the $M_2$-assisted reconstruction to compute observables sensitive to the CP-phase at the $t\bar{t}$ rest frame, where the new physics sensitivity can be enhanced.
Performing a side-band analysis at the LHC to control the continuum $t\bar{t}b\bar{b}$ background, we find that the Higgs-top strength and CP-phase can be probed up to $\delta\kappa_t\lesssim 20\%$ and $| \alpha | \lesssim 36^\circ$  at 95\%~CL, respectively. We also derive that a similar analysis at a 100~TeV future collider could further improve the precision to $\delta\kappa_t\lesssim 1\%$ and $| \alpha| \lesssim 1.5^\circ$, where the CP-odd observables play a crucial role, boosting the sensitivity on the CP-phase.
}


\begin{document}

\maketitle
\flushbottom

\section{Introduction}
\label{sec:intro}
The possible existence of new CP-violating interactions can play a significant role in explaining the matter-antimatter asymmetry of the universe~\cite{Sakharov:1967dj}. Thus, the search for new sources of CP-violation is a clear target in the quest for new physics. A prominent path in this program  is  to boost our current understanding of the Higgs boson couplings. Remarkably, from the theoretical viewpoint, there are some couplings more susceptible to display larger new physics effects than others. For instance, the well studied CP-odd Higgs-vector boson interactions can appear only through operators of dimension-6 or higher~\cite{Buchmuller:1985jz,Grzadkowski:2010es}, being naturally suppressed by the new physics scale. Nevertheless, the CP-odd Higgs-fermion couplings can manifest already at the tree level~\cite{Buckley:2015vsa}, granting naturally sizable CP violation effects.

Owning to its magnitude,  the top quark  Yukawa coupling is central to this discussion and could be most sensitive to physics  beyond the Standard Model (SM). While it is possible to probe this interaction with loop-induced processes~\cite{Brod:2013cka,Dolan:2014upa,Englert:2012xt,Bernlochner:2018opw,Englert:2019xhk,Gritsan:2020pib,Bahl:2020wee}, the \emph{direct} measurement via $pp\to t\bar{t}h$ production plays a crucial role, disentangling possible new physics contributions~\cite{Ellis:2013yxa,Boudjema:2015nda,Buckley:2015vsa,Buckley:2015ctj,Gritsan:2016hjl,Goncalves:2016qhh,AmorDosSantos:2017ayi,Azevedo:2017qiz,Goncalves:2018agy,ATLAS:2018mme,CMS:2018uxb,Bortolato:2020zcg,MammenAbraham:2021ssc}. Recently ATLAS and CMS collaborations have reported the first experimental CP analyses using the $t\bar{t}h$ production~\cite{ATLAS:2020ior,CMS:2020cga}.
These initial studies focus solely on the di-photon final state, $h\to \gamma \gamma$. ATLAS excludes the CP-mixing angle greater than $43^\circ$  and CMS above $55^\circ$ at 95\% confidence level (CL).

The high luminosity LHC (HL-LHC) projections, performed by both ATLAS and CMS, indicate that the $t\bar{t}h$ channel in the di-photon final state will result in dominant sensitivities for new physics searches, with a signal strength limit of $\delta\mu_{tth}^{\gamma\gamma}\lesssim 5.9\%$ at 68\%~CL~\cite{Cepeda:2019klc}. Despite the limited signal statistics, the di-photon channel highly benefits from controlled backgrounds through a side-band analysis. At the same time, the dominant Higgs decay to bottom quarks, ${\mathcal{BR}(h\to b\bar{b})\sim 58\%}$, will only grant sub-leading limits, $\delta\mu_{tth}^{bb}\lesssim 10.7\%$, as its search endures a substantial QCD background associated with sizable uncertainties~\cite{ATLAS:2017fak,CMS:2018hnq}.

Focusing on the 100~TeV Future Circular Collider~(FCC) and semi-leptonic top pair final states,  Ref.~\cite{Plehn:2015cta} shows that a combination of side-bands and $t\bar{t}h/t\bar{t}Z$ ratios can uplift the top-quark Yukawa strength determination with the $t\bar{t}(h\to b\bar{b})$ channel. Inspired by  this finding, we apply a similar methodology to control the background uncertainties for both the 14~TeV HL-LHC and 100~TeV FCC,  deriving the top Yukawa CP-phase sensitivity. Instead of the semi-leptonic top pair final states, we consider the di-leptonic mode. Besides the significant background suppression, this final state benefits from the larger top quark spin analyzing power associated with charged leptons~\cite{Bernreuther:2010ny}, resulting in stronger probes to the CP violation through spin correlations.

This paper is organized as follows. In Section~\ref{sec:theory}, we  present the theoretical setup and discuss the relevant observables sensitive to the Higgs-top CP-phase. In Section~\ref{sec:rec}, we review the adopted  reconstruction method for the di-leptonic top pair final state, which is relevant to build up  prominent observables sensitive to new physics. We then move on to a detailed analysis in Section~\ref{sec:ana}, where we derive the projected sensitivities to the CP-phase from the HL-LHC and FCC, exploring the side-bands and the correlation between the $t\bar{t}h$ and $t\bar{t}Z$ uncertainties. Finally, we present a summary in Section~\ref{sec:summary}.

\section{Theoretical Setup}
\label{sec:theory}
We parameterize the Higgs-top interaction as
\begin{align}
    \mathcal{L} \supset -\frac{m_t}{v}\kappa_t \bar{t}\left(\cos\alpha+i\gamma_5\sin\alpha\right)  t \, h\,,
\end{align}
where $\kappa_t$ is a real number that modulates the interaction strength, $\alpha$ is the CP-phase, and ${v=246}$~GeV is the SM Higgs vacuum expectation value. The SM hypothesis displays $\kappa_t=1$ and $\alpha=0$. In contrast, a purely CP-odd particle would have $\alpha=\pi/2$.

Among the several probes sensitive to the CP-structure of the Higgs-top interaction in the $t\bar{t}h$ channel, the observables defined in the $t\bar{t}$ center-of-mass frame play a special role~\cite{Goncalves:2018agy}. First, this reference frame allows the definition of phenomenologically relevant CP-odd observables arising from fully anti-symmetric tensor products. A prominent example of this sort is the tensor product involving the two top quarks and the two final state charged leptons, that carry maximal top spin analyzing power, $\epsilon_{\mu\nu\rho\sigma} p_{t}^\mu p_{\bar{t}}^\nu p_{\ell^+}^\rho p_{\ell^-}^\sigma$. This initially complex phenomenological probe can be opportunely simplified in the top pair frame to a simple triple product, $\vec p_t \cdot  (\vec p_{\ell^+}\times \vec p_{\ell^-})$, being more suitable for collider studies. In particular, this mathematical property can be used to define the angular correlation between the charged leptons in the $t\bar{t}$ rest frame~\cite{Goncalves:2018agy}
\begin{equation}
\Delta \phi_{\ell\ell}^{t\bar t} =
    \text{sgn} \left[\vec{p}_{t} \cdot (\vec{p}_{\ell^+} \times \vec{p}_{\ell^-})\right]
     \arccos \left[ \frac{\vec{p}_{t} \times \vec{p}_{\ell^+}}{|\vec{p}_{t} \times \vec{p}_{\ell^+}|} \cdot \frac{\vec{p}_{t} \times \vec{p}_{\ell^-}}{|\vec{p}_{t} \times \vec{p}_{\ell^-}|}\right]\,,
\end{equation}
which is also sensitive to the sign of the CP-phase. Second, there are additional robust observables that also display relevant sensitivity in this frame, such as the Collins-Soper angle $\theta^*$. The $\theta^*$ observable is the production angle of the top with respect to the beam axis in the $t\bar{t}$ center-of-mass frame.

While the definition of several phenomenological probes in the top pair center-of-mass frame is a desirable ingredient to uplift the sensitivity to the CP-phase, the presence of two neutrinos in the di-leptonic $t\bar{t}h$ channel brings a challenge for the event reconstruction in a hadron collider environment. In the next section, we will describe a mass minimization method to efficiently overcome this obstacle. This approach has been proven robust against parton shower, hadronization, and detector effects~\cite{Goncalves:2018agy}.

\section{Brief Review on the $M_2$-Assisted Reconstruction of Top Pair}
\label{sec:rec}
\begin{figure}[t!]
\centering
\includegraphics[scale=1.1]{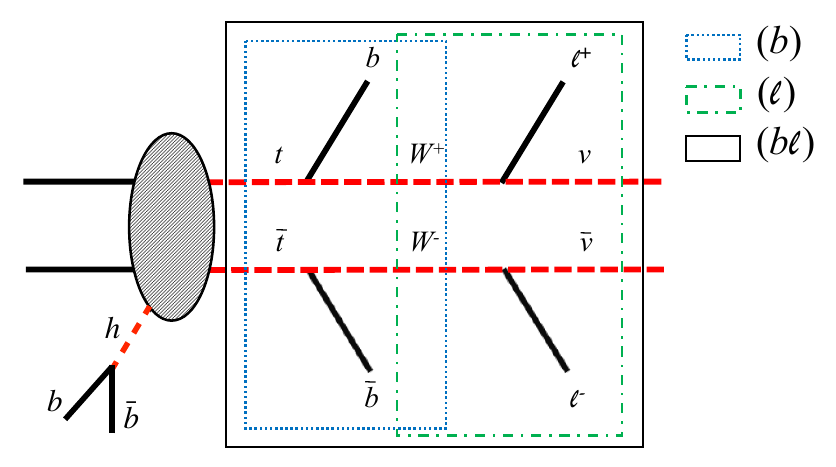}
\caption{The event topology considered in this paper. The blue dotted, the green dot-dashed, and the black solid boxes indicate the subsystems $(b)$, $(\ell)$, and $(b\ell)$, respectively. \label{fig:decaysubsystem}}
\end{figure}

The event topology considered in this study is depicted in~\autoref{fig:decaysubsystem}, where the blue dotted, the green dot-dashed, and the black solid boxes indicate the three subsystems $(b)$, $(\ell)$, and $(b\ell)$, respectively~\cite{Burns:2008va}. The Higgs decays to a pair of bottom quarks and the associated top quarks both decay leptonically.
For such events with two missing particles, the on-shell constrained $M_2$ variable provides a good estimation for the unobserved invisible momenta and thus can be useful to discriminate the combinatorial ambiguities~\cite{Barr:2011xt,Debnath:2017ktz,Kim:2017awi}. The $M_2$~\cite{Barr:2011xt} is defined as a $(3+1)$-dimensional version of the \emph{stransverse} mass $M_{T2}$~\cite{Lester:1999tx}:
\begin{align}
M_{2} (\tilde m) &\equiv \min_{\vec{q}_{1},\vec{q}_{2}}\left\{\max\left[M_{P_1}(\vec{q}_{1},\tilde m),\;M_{P_2} (\vec{q}_{2},\tilde m)\right] \right\} , \label{eq:m2def}
\\
\mpt &=\vec{q}_{1T}+\vec{q}_{2T}  \;,\nonumber
\end{align}
where the {\em actual} parent masses, $M_{P_i}$ ($i=1, \, 2$), are considered in the minimization instead of their transverse masses, as is done in $M_{T2}$.
Note that the minimization is performed over the 3-component momentum vectors $\vec{q}_{1}$ and $\vec{q}_{2}$ of the two missing particles~\cite{Barr:2011xt} under the missing transverse momentum constraint, $\mpt=\vec{q}_{1T}+\vec{q}_{2T}$.
We use the zero test mass ($\tilde m=0$), as two missing particles are neutrinos in our study.
At this point $M_{T2}$ and $M_2$ are known to be equivalent, in the sense that the resulting two variables lead to the same numerical value, $M_{T2} = M_2 \leqslant \max (M_{P_1} ,M_{P_2})$~\cite{Ross:2007rm,Barr:2011xt,Cho:2014naa}.

However, $M_2$ provides more flexibility in incorporating additional kinematic constraints. For example, in the $t \bar t$-like production considered in this paper ($t\bar t + X$, where the transverse momentum of $X$ is known), we could use the experimentally measured $W$-boson mass, $m_W$, and introduce the following variable in the ($b\ell$) subsystem:
\bea
M_{2CW}^{(b\ell)} (\tilde m) &\equiv& \min_{\vec{q}_{1},\vec{q}_{2}}\left\{\max\left[M_{t_1}(\vec{q}_{1},\tilde m),\;M_{t_2} (\vec{q}_{2},\tilde m)\right] \right\}, \label{eq:m2CWdef}\\
\mpt&=&\vec{q}_{1T}+\vec{q}_{2T}   \,,\nonumber  \\
M_{t_1}&=& M_{t_2}\,, \nonumber  \\
M_{W_1}&=& M_{W_2} = m_W \,.\nonumber
\eea
Here, the second constraint $M_{t_1}= M_{t_2}$ requires the equality of two parent mass without use of a specific numerical value, while the true $W$ mass is used in the third constraint $M_{W_1}= M_{W_2} = m_W$.
Similarly, taking the top quark mass $m_t$ in the minimization, we can define a new variable in the ($\ell$) subsystem:
\bea
M_{2Ct}^{(\ell)} (\tilde m) &\equiv& \min_{\vec{q}_{1},\vec{q}_{2}}\left\{\max\left[M_{W_1}(\vec{q}_{1},\tilde m),\;M_{W_2} (\vec{q}_{2},\tilde m)\right] \right\},\label{eq:m2Ctdef}\\
\mpt&=&\vec{q}_{1T}+\vec{q}_{2T}   \,, \nonumber  \\
M_{W_1}&=& M_{W_2} \,, \nonumber  \\
M_{t_1}&=& M_{t_2} = m_t \,.\nonumber
\eea
These distributions exhibit sharper end point structure in their kinematic distribution, due to additional mass information in the minimization, {\it i.e.,} $M_2^{(b\ell)} \leqslant M_{2CW}^{(b\ell)} \leqslant m_t $ and $M_2^{(\ell)} \leqslant M_{2Ct}^{(\ell)} \leqslant m_W $~\cite{Barr:2011xt,Cho:2014naa}.

While these mass-constraining variables are proposed for mass measurement originally,
one could use them for other purposes, such as measurement of spins and couplings~\cite{Baringer:2011nh,Debnath:2017ktz}. In our study, we use these variables to fully reconstruct the final state of our interest, with the unknown neutrino momenta obtained via minimization procedure. These momenta may or may not be true momenta of the missing neutrinos, but they provide important non-trivial correlations with other visible particles in the final state, which improves the reconstruction.

Based on Ref.~\cite{Debnath:2017ktz}, we take advantage of the kinematic features of the following 3-dimensional mass space:
\begin{equation}
\Big(   \,
m_{b\ell}^{max}-\max_i\{m^{(i)}_{b\ell}\}, \,
m_t - M_{2CW}^{(b\ell)}, \,
m_W - M_{2Ct}^{(\ell)} \,
\Big) ,
\label{setofthreeWt}
\end{equation}
where $m^{(i)}_{b\ell}$ is the invariant mass of $b$ and $\ell$ in the $i$-th side ($i=1,2$), and $m_{b\ell}^{max} = \sqrt{ m_t^2 - m_W^2}$ (in the $m_b \to 0$ limit). Since there are two possible ways of paring $b$ and $\ell$ in the di-lepton channel of the $t \bar t$-like events, we repeat the same calculation for each partition. The correct combination would respect the anticipated end points of $m_{b\ell}$, $M_{2CW}^{(b\ell)}$, and $M_{2Ct}^{(\ell)}$, leading to positive values of three components in above 3-dimensional mass space. On the other hand, the wrong pairing could give either sign. By requiring that the partition which gives more ``plus" sign as the ``correct" one, we can resolve the two-fold ambiguity. Then, we treat the corresponding momenta of two missing particles, which are obtained via the minimization procedure, as ``real'' momenta of two missing neutrinos. If both partitions give the same numbers of positive and negative signs, we discard such events, since they are  ``unresolved cases''.
We note that we assign the negative sign for a partition, if a viable solution is not found during minimization. This is because the wrong pairing would fail more often than the correct one.

From Ref.~\cite{Debnath:2017ktz}, the efficiency of this method at the parton-level is known to be about 88\%, including unresolved events with a coin flip, 50\% probability of picking the right combination. Since we ignore those events to obtain a high-purity sample, the corresponding efficiency becomes 83\%. In our analysis, we find that the final efficiency is about 78\% including more realistic effects such as parton-shower and hadronization.

We use \verb|OPTIMASS|~\cite{Cho:2015laa} for the minimization to obtain momenta of two invisible neutrinos, following the reconstruction method described above.
With the obtained neutrino momenta, now we can reconstruct momenta of $W$-bosons and top quarks for the measurement of the CP-phase. The fully reconstructed top quark momenta allow the Lorenz-boost transformation from the lab frame to the $t\bar t$ rest frame, which is crucial for our analysis.

\section{Analysis}
\label{sec:ana}
To directly probe the Higgs-top CP-structure, we explore the $pp\rightarrow t\bar{t}h$ production with the Higgs boson decay to bottom quarks,  $h\rightarrow b\bar{b}$, associated with di-leptonic top quarks. We derive the new physics sensitivity for both the 14~TeV HL-LHC and 100~TeV FCC~\cite{Abada:2019lih}. Our signal is characterized by two opposite sign charged leptons, $\ell=e$ or $\mu$, and four $b$-tagged jets. The major backgrounds, in order of relevance, are $t\bar{t}b\bar{b}$ and $t\bar{t}Z$. The signal and background event samples are simulated with {\tt MadGraph5\_aMC@NLO}~\cite{Alwall:2014hca}. To include higher order effects, we rescale the $t\bar{t}h$, $t \bar t b \bar b$, and $t \bar t Z$ cross-sections with flat next-to-leading order k-factors derived with {\tt MadGraph5\_aMC@NLO}. The parton shower, hadronization, and underlying event effects are included with {\tt Pythia6}~\cite{Sjostrand:2006za}. To secure top-quark spin-correlation effects, the top-quark decays are performed with {\tt MadSpin}~\cite{Artoisenet:2012st}.

The adopted analysis strategy explores the boosted Higgs regime. Along with the background suppression~\cite{Butterworth:2008iy,Plehn:2009rk}, this kinematic configuration opportunely enhances the top-quark spin correlation effects~\cite{Buckley:2015vsa}. We begin our analysis requiring two isolated and opposite charged leptons with $p_{T\ell}>20$~GeV and $|\eta_\ell|<2.5$. The hadronic part of the event is first reclustered using the Cambridge/Aachen jet algorithm with $R=1.2$, requiring one or more boosted fat-jets with $p_{TJ}>200$~GeV and $|\eta_J|<2.5$. The jet reclustering is performed with {\tt FastJet}~\cite{Cacciari:2011ma}. We demand one of the fat-jets to be Higgs-tagged via the BDRS algorithm~\cite{Butterworth:2008iy}, imposing that its two hardest subjets are $b$-tagged. Since the complete analysis displays four $b$-tags, we take advantage of the improvements reported by ATLAS, associated with the central tracking system for the operation at the HL-LHC, and use a work point with a large $b$-tagging efficiency~\cite{CERN-LHCC-2017-021}. We assume 85\% $b$-tagging efficiency associated with 1\% (25\%) mistag rate for light-jets ($c$-jets), being consistent with the experimental studies from the ATLAS collaboration.

Since the signal event does not display another hadronic heavy particle decay, we can safely suppress the underlying event contamination by using a smaller jet size for the rest of the event. Hence, after the Higgs tagging, we remove the fat-jet associated with the Higgs boson and recluster the remaining hadronic activity with the anti-k$_t$ jet algorithm with $R=0.4$, $p_{Tj}>30$~GeV, and $|\eta_j|<2.5$. We demand two extra $b$-tagged jets to control possible extra backgrounds. More details on the event selections are described in~\autoref{tab:Cutflow1} and~\autoref{tab:Cutflow100tev} for the 14~TeV HL-LHC and 100~TeV FCC, respectively.

\subsection{CP measurement at 14~TeV HL-LHC}

\begin{table*}[t]
\begin{center}
\setlength{\tabcolsep}{0.9mm}
\renewcommand{\arraystretch}{1.4}
\scalebox{1.0}{
\hspace*{-20pt}
\begin{tabular}{|c||c|c|c||c|c|}
\hline
    cuts                                                                                                     & $t \bar{t} h $   & $t \bar{t} b \bar{b} $           & $t \bar{t} Z $         & $\sigma$        \\  \hline
    \hline
~~$N_{h}= 1$, $4b$-tags, $p_T^\ell  > 20\ {\rm GeV}$, $|\eta^\ell | < 2.5$~~  &  \multirow{2}{*}{ ~$0.358$~ }      &   \multirow{2}{*}{ ~4.08~ }         &  \multirow{2}{*}{~0.106~}  &  \multirow{2}{*}{ ~9.47~ }       \\
$p_T^j  > 30\ {\rm GeV}$, $|\eta^j| < 2.5$, $N_{j} \geqslant 2$, $N_{\ell} = 2$    &                                               &                                            &                                        &                 \\  \hline
$50~\ {\rm GeV} < m_J^\text{BDRS} < 150\ {\rm GeV}$                                                                       & $0.306$                               & $2.18$                                 & $0.0971$            &     $10.9$  \\  \hline
Resolving combinatorics                                                                       & $0.239$                               & $1.47$                               & $0.0796$                       &    $10.3$    \\  \hline 
\end{tabular}}
\end{center}
\caption{Cumulative cut-flow table showing  cross-section in fb for  $t\bar{t}h$ signal ($\kappa_t=1,\alpha = 0$) and leading backgrounds $t\bar{t}b\bar{b}$ and $t\bar{t}Z$ at the 14~TeV LHC. Significances ($\sigma$) are calculated for a luminosity of 3~$\rm{ab}^{-1}$.  We apply a flat $b$-tag rate of $\epsilon_{b \rightarrow b} = 0.85$ and a light-jet ($c$-jet) mistag rate of $\epsilon_{j \rightarrow b} = 0.01$ ($\epsilon_{c \rightarrow b} = 0.25$) in accordance with the improvements reported by the ATLAS collaboration for the HL-LHC performance~\cite{CERN-LHCC-2017-021}.
}
\label{tab:Cutflow1}
\end{table*}

\begin{figure}[b]
    \centering
    \includegraphics[width=0.5\textwidth]{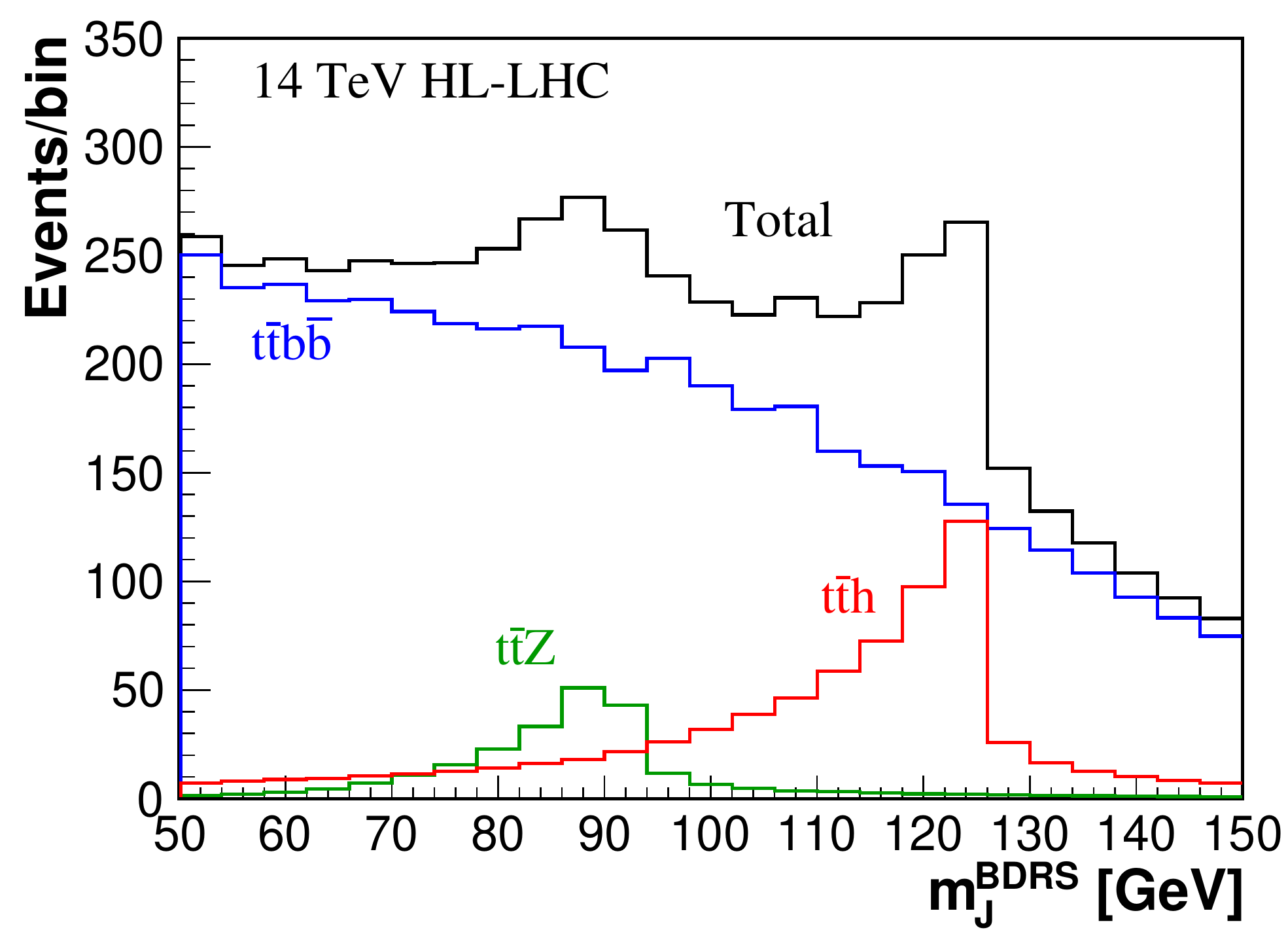}
    \caption{
    The invariant mass distributions of signal and background for the BDRS tagged fat-jet $m_J^{\rm BDRS}$ at the 14~TeV HL-LHC. We show $t\bar th$  signal (red), the $t\bar{t}b\bar{b}$ (blue), and $t\bar tZ$ (green) in a non-stacked format. The full stacked result is also presented (black). We assume 3~ab$^{-1}$ of integrated luminosity. For more details on the cut-flow analysis, see~\autoref{tab:Cutflow1}.
    \label{fig:mj}}
\end{figure}

\begin{figure*}[t]
\begin{center}
\includegraphics[width=0.33\textwidth,clip]{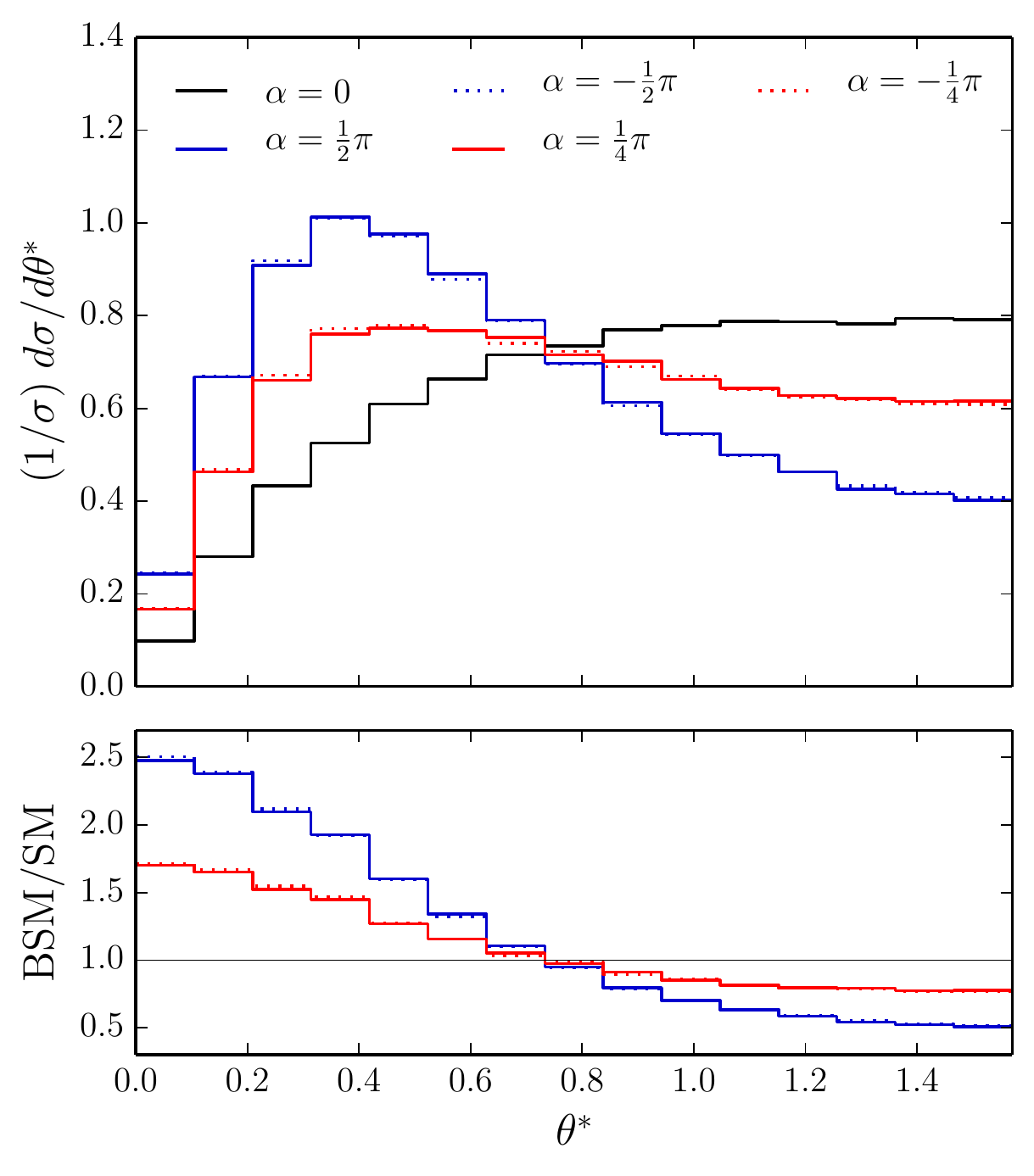} \hspace*{-0.17cm}
\includegraphics[width=0.33\textwidth,clip]{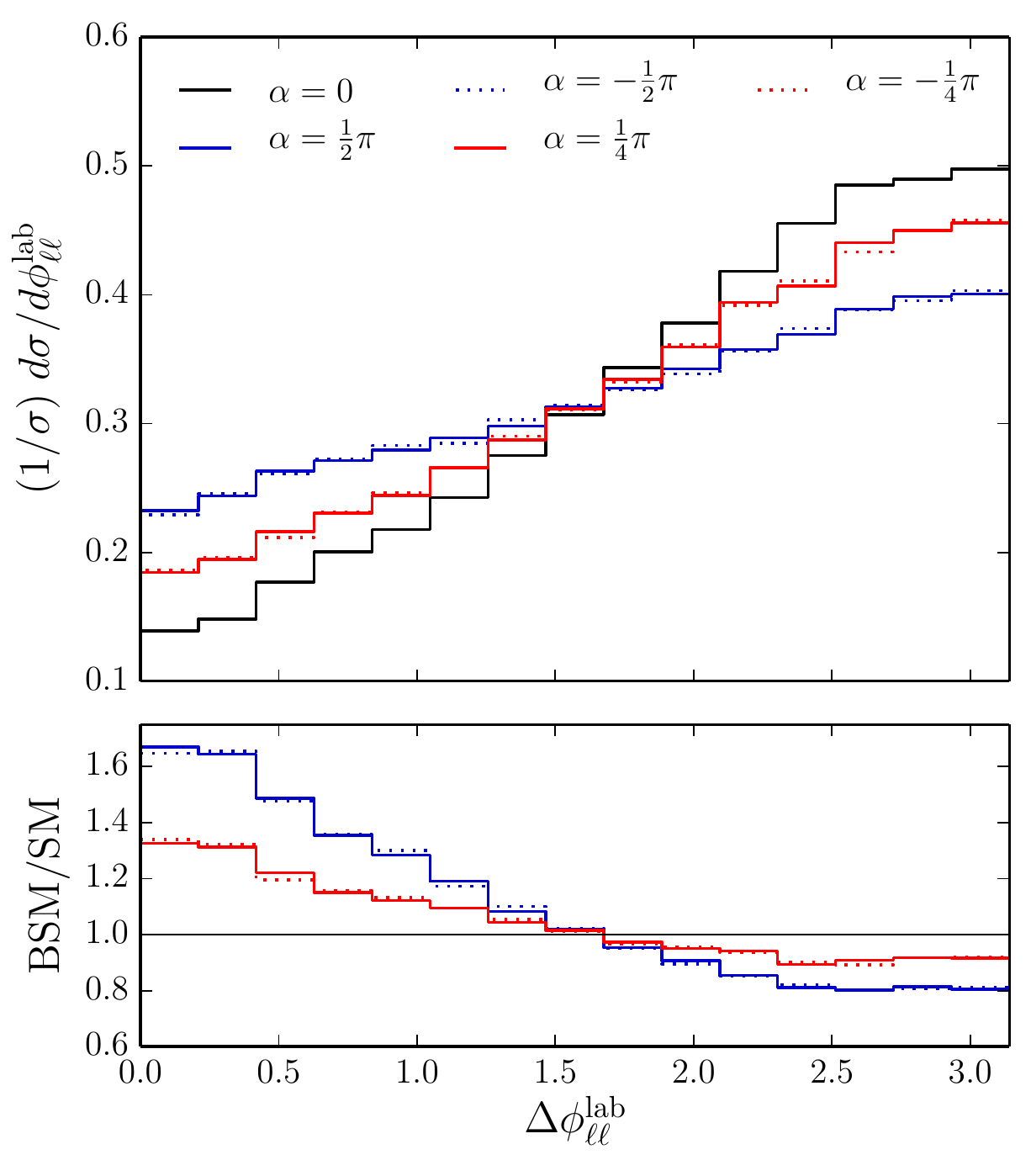} \hspace*{-0.17cm}
\includegraphics[width=0.33\textwidth,clip]{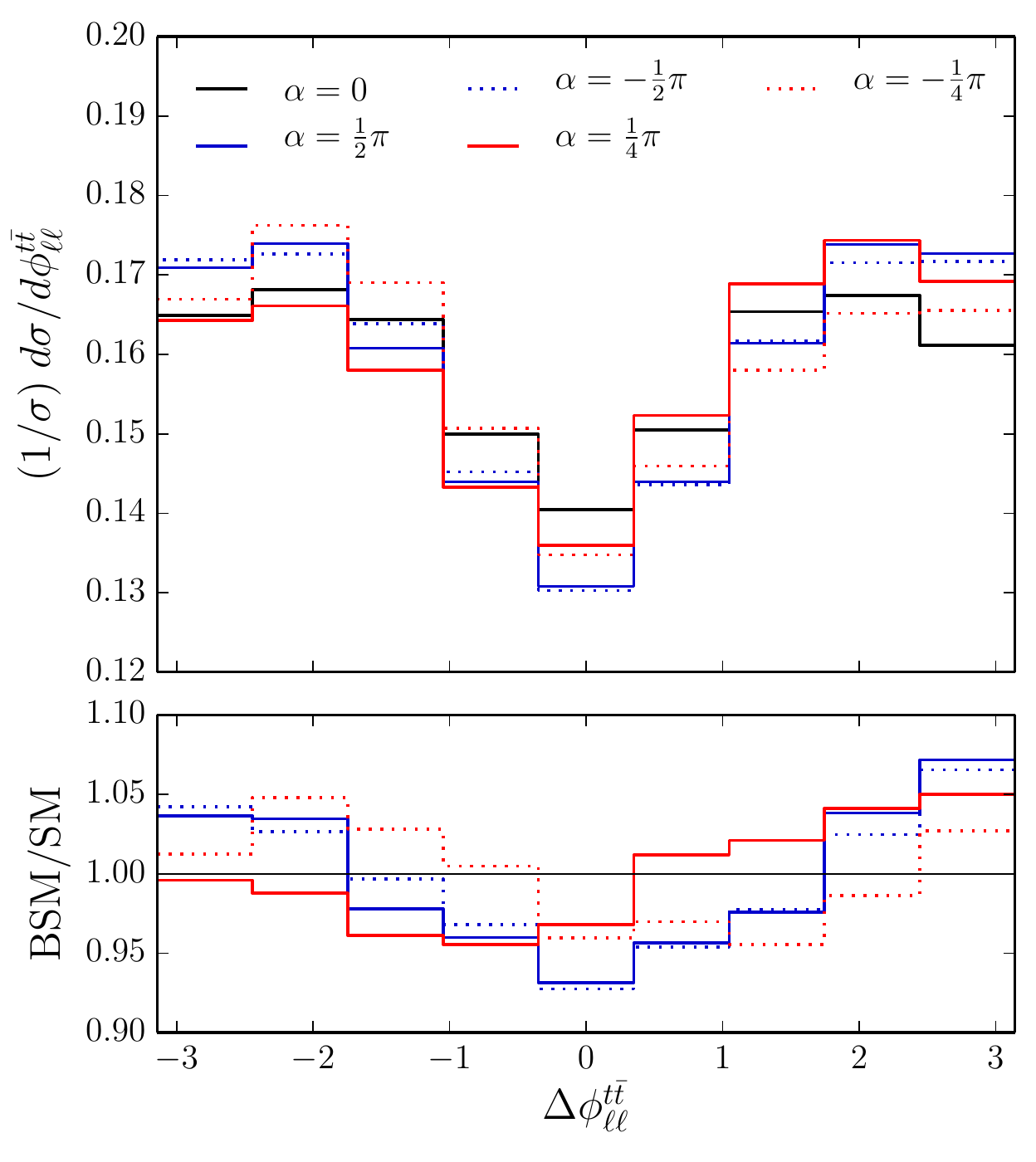}
\caption{$\theta^\ast$ (left), $\Delta \phi^\text{lab}_{\ell \ell }$ (middle),  and $\Delta \phi^{t \bar{t}}_{\ell \ell }$ (right) distributions for the $t\bar{t}h$ samples at the 14~TeV LHC, after the basic cuts described in~\autoref{tab:Cutflow1}, and resolving the combinatorial problem.
}\label{ttg_FR}
\end{center}
\end{figure*}

We present in~\autoref{fig:mj} the invariant mass distribution of the BDRS tagged fat-jet, $m_J^\text{BDRS}$, for the signal and background samples at the 14~TeV HL-LHC with 3~ab$^{-1}$ of data. Remarkably, the Higgs signal from the $t\bar{t}h$ sample results in a clear peak structure around the Higgs mass. In particular, this is a result of the BDRS filtering that promotes the invariant mass associated with the fat-jet to a robust observable efficiently controlling the pile-up effects~\cite{ATLAS:2012am}.

We show in~\autoref{ttg_FR} the relevant CP sensitive probes for the $t\bar{t}h$ samples used in this analysis, namely $\theta^\ast$ (left), $\Delta \phi^\text{lab}_{\ell \ell }$ (middle), and $\Delta \phi^{t \bar{t}}_{\ell \ell }$ (right). In the bottom of each panel, we show the ratio of non-zero CP phase to the SM prediction ($\alpha=0$).
These distributions are presented after  reconstruction of the top quark pair, with the selections outlined in~\autoref{tab:Cutflow1}.
The $\Delta \phi^{t \bar{t}}_{\ell \ell }$ distribution exhibits the sensitivity on the sign of the CP-phase, while both
$\theta^\ast$  and $\Delta \phi^\text{lab}_{\ell \ell }$ are CP-even variables.
We observe that the $t\bar{t}$ reconstruction described in~\autoref{sec:rec} is robust, resulting in observables with strong modulations for distinct top Yukawa CP-phases even at the hadron level.

\begin{figure*}[!t]
    \centering
    \includegraphics[width=\textwidth]{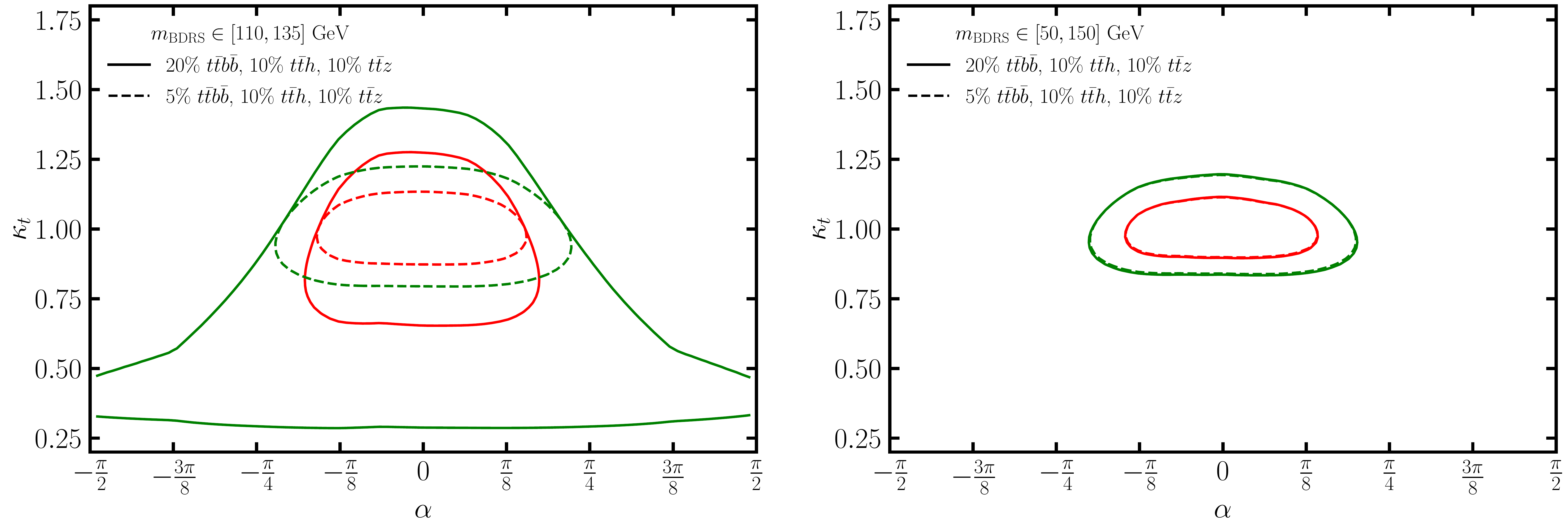}
    \caption{The exclusion at 68\% (red) and 95\% (green) CL in the $\alpha$-$\kappa_t$ plane at the 14~TeV LHC with $3\,{\rm ab}^{-1}$ for a narrow (left) and wide (right) mass window.
    20\% systematics (5\%) for $t \bar t b \bar b$ is assumed in solid (dotted) curves, while 10\% systematics is used for both $t\bar t Z$ and $t \bar th$.
    }
    \label{fig:alpha_kt_14TeV}
\end{figure*}

To enhance the signal sensitivity, we perform a binned log-likelihood analysis exploring the Higgs candidate invariant mass profile, in the signal range ${m_J^\text{BDRS}\in [110,135]\,\rm GeV}$, together with the CP-sensitive observable $\theta^*$, defined at the $t\bar{t}$ center-of-mass frame.  Since the considered $t\bar{t}h$ channel with $h\to b\bar{b}$ typically confronts a large $t\bar{t}b\bar{b}$ background, which has a significant uncertainty~\cite{ATLAS:2017fak,CMS:2018hnq}, the  final result displays relevant correlation with the considered background uncertainties. To estimate this effect, we derive the new physics sensitivity on the $(\alpha,\kappa_t)$ plane for two scenarios. In the first case, we assume that  $t\bar{t}b\bar{b}$ background rate has 20\% of uncertainty, which is included as a nuisance parameter. The magnitude of the considered error is similar to the current experimental analyses~\cite{ATLAS:2017fak,CMS:2018hnq}. For the second case, we assume an optimistic scenario with 5\% error. The uncertainties on the $t\bar{t}h$ and $t\bar{t}Z$ samples are assumed to be 10\% for both scenarios~\cite{Plehn:2015cta}. The result of this analysis is presented in the left panel of~\autoref{fig:alpha_kt_14TeV}.
We obtain that the CP-mixing angle can be constrained to $|\alpha|\lesssim 32^\circ$ at 68\%~CL at the HL-LHC for both scenarios. At the same time, we find that the sensitivity from $\kappa_t$ to the systematic error is more pronounced. While in the first scenario we can constrain the top Yukawa strength to $\delta\kappa_t \lesssim 0.3$, the more optimistic case leads to  $\delta\kappa_t\lesssim 0.15$.

We note that in the absence of the shape information of the $\theta^*$ distribution, there exists a flat direction in the $(\alpha,\kappa_t)$ plane, irrespective of the considered uncertainty scenarios, where two red (or green) curves meet tangentially as shown in the left panel of~\autoref{fig:alpha_kt_14TeV}.
In other words, along the flat direction, there is no constraint on the values of $\kappa_t$ and $\alpha$. The constraint stems from the shape of $\theta^*$ distribution. Therefore, along that flat direction, the limits on $(\alpha,\kappa_t)$ will not change and the considered uncertainties of $t\bar tb \bar b$ do not affect the fit.
Further, when $\kappa_t\sim 0.4$, there is no sensitivity on $\alpha$ for the case with large systematics (20\% for $t\bar t b\bar b$). This is because the signal rate is suppressed for $\kappa_t$ around that region, and thus we gain no information from $\theta^*$ distribution, while the large systematics from the $t\bar tb\bar b$ can compensate the total event rate. When $\kappa_t$ is even smaller, the signal rate is further suppressed, and the fluctuation from background alone cannot explain the total event rate, excluding the small $\kappa_t$ region.

\begin{figure*}[t]
	\centering
	\includegraphics[width=0.5\textwidth]{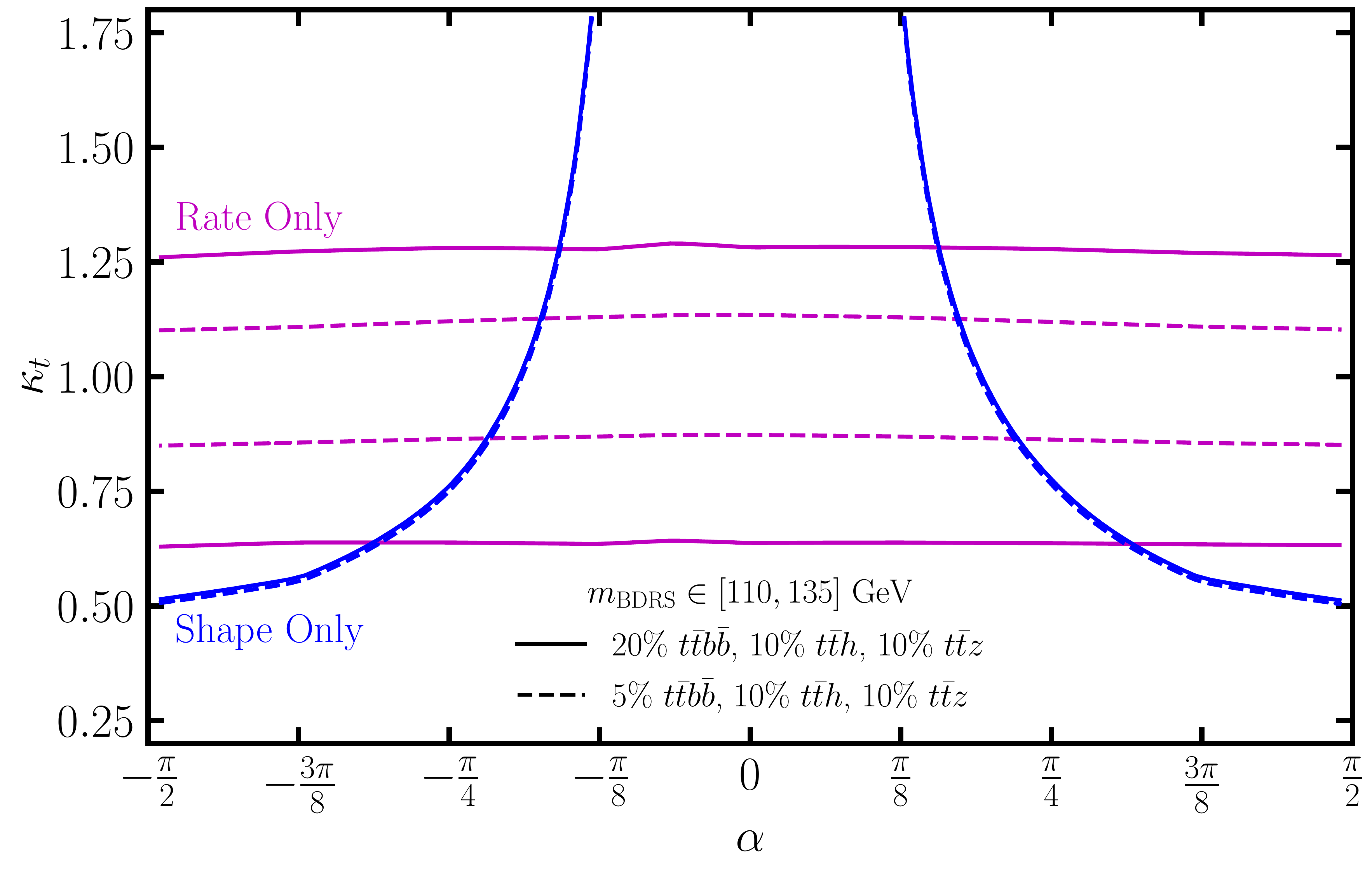}
	\caption{The individual exclusion at 68\%~CL from the rate-only measurement (magenta) and shape-only measurement (blue) for two different assumptions on the systematic uncertainty for $t\bar t b\bar b$, 20\% in solid and 5\% in dotted.
	\label{fig:individualfit}}
\end{figure*}

This observation in the left panel of~\autoref{fig:alpha_kt_14TeV} can be more clearly understood by studying the separate exclusions. In~\autoref{fig:individualfit}, we show the individual exclusion at 68\%~CL from the rate-only  measurement (in magenta) and the shape-only measurement (in blue). The exclusion with the rate-only does not have significant sensitivity on $\alpha$, since the signal rate remains roughly fixed as the CP angle $\alpha$ varies. This arises from a combination of two effects that approximately cancel out. While the inclusive $t\bar t h$ production cross-section decreases when scanning $\alpha$ from $0$ to $\pi/2$~\cite{Demartin:2015uha}, the signal acceptance increases for larger $\alpha$ due to a sizable difference in kinematics between CP-even ($\alpha=0$) and CP-odd ($\alpha=\pi/2$) in the boosted regime~\cite{Goncalves:2018agy}. The two factors roughly cancel, leading to suppressed differences in event rate for distinct $\alpha$. On the other hand, the shape-only exclusion exhibits sensitivity on $\alpha$. Hence, one can recover the general profile of the exclusion in~\autoref{fig:alpha_kt_14TeV} by combining the four curves in~\autoref{fig:individualfit}.

To illustrate how to reduce the systematics for $t\bar t b\bar b$ in a realistic measurement, we enlarge the mass range of the Higgs candidate to ${m_J^{\rm BDRS}\in [50,150]\,\rm GeV}$. In this way, the events outside the Higgs peak, which mainly come from $t\bar t b\bar b$, can be used together with the shape of $m_J^{\rm BDRS}$ distribution of $t\bar tb\bar b$ from MC simulation within the binned log-likelihood method. By fitting to a broader range of $m_J^{\rm BDRS}$, we have a better control of the uncertainties of $t\bar tb\bar b$.
The results are shown in the right panel of~\autoref{fig:alpha_kt_14TeV}. We find that this analysis depletes the influence of the systematic uncertainties, leading to similar results for the two considered systematic uncertainty scenarios. The obtained limits are $|\alpha|\lesssim26^\circ~(36^\circ)$ and $\delta\kappa_t\lesssim 0.12~(0.2)$ at 68\%~(95\%) CL. Using the wider mass window, the log-likelihood analysis takes full advantage of the shape information of $t\bar t h$ and $t \bar t b \bar b$ events.

It is illuminating to analyze the number of signal and background events in the Higgs peak and side-bands to infer the uncertainty suppression.
Around the Higgs peak $m_J^{\rm BDRS}\in [110,135]~\rm GeV$, the number of signal events at the HL-LHC is $N_{tth}=N_S=N_{\rm total}-N_{ttbb}\approx 402$, whereas the number of background events is $N_B=N_{ttbb}\approx 865$. For this discussion, given the considered invariant mass window for $m_J^{\rm BDRS}$, the $t\bar tZ$ background makes a subleading contribution.  Assuming 20\% systematics ($\Delta N_{ttbb} = 0.2\times N_{ttbb}$) for the $t\bar t b \bar b$ events, one can estimate the uncertainty, $\Delta N_S$, for $N_S$ as follows,
\begin{align}
	(\Delta N_S)^2 &= \big (\sqrt{N_{\rm total}} \big )^2 + \big (\Delta N_{ttbb} \big )^2\nonumber\\
	&=N_{\rm total} + \big (0.2\times N_{ttbb} \big )^2\nonumber\\
	&\approx 31,196\nonumber\\
	&\approx \big (0.44\times N_S \big )^2 \, ,
\end{align}
which is very large due to the background events in the signal region.
However, if we use the side-band events to estimate the $t\bar tb\bar b$ events within the Higgs peak, we can suppress the uncertainties in the signal region.
For $m_J^{\rm BDRS}\in [50,110] \oplus [135, 150]~\rm GeV$, we have $N_{\rm sideband}\approx3,543$.
Then, the signal ($t\bar th$) events can be estimated directly from measurement as $N_S = N_{\rm total} - \kappa N_{\rm sideband}$, where $N_{\rm total}=1,267$ for $m_J^{\rm BDRS}\in[110,135]~\rm GeV$. Here, $\kappa$ is the ratio of the number of background events in the signal region ($N_B=865$) to that in the side-band region ($N_{\rm sideband}=$3,543), which can be estimated from MC or from the shape of the inferred background distribution using the side-bands. As an illustration, we fix $\kappa$ as $\kappa=N_B/N_{\rm sideband}$. Then the uncertainties for $N_S$ is calculated as
\begin{align}
	(\Delta N_S)^2 &= \big (\sqrt{N_{\rm total}} \big )^2 + \big (\kappa \Delta N_{\rm sideband} \big )^2 \nonumber \\
	&= N_{\rm total} + \big (\kappa \sqrt{N_{\rm sideband}} \big )^2 \nonumber \\
	&= N_{\rm total} + \frac{N_B^2}{N_{\rm sideband}}\nonumber\\
	&\approx 1,478.2\nonumber\\
	& \approx  \big (0.096\times N_S \big )^2 \, ,
\end{align}
which has significantly lower uncertainties for $N_S$ with the aid of the events from the side-bands~\cite{Plehn:2015cta}.
Similar background control regions are actively used in experimental analyses, for example, for the
$h \to \gamma \gamma$ channel~\cite{CMS:2012qbp}, and $Wh / Zh$ production in the $h \to b \bar b$ decay channel~\cite{ATLAS:2020fcp}.

\subsection{CP measurement at 100~TeV FCC-hh}

\begin{figure}[b!]
    \centering
    \includegraphics[width=0.5\textwidth]{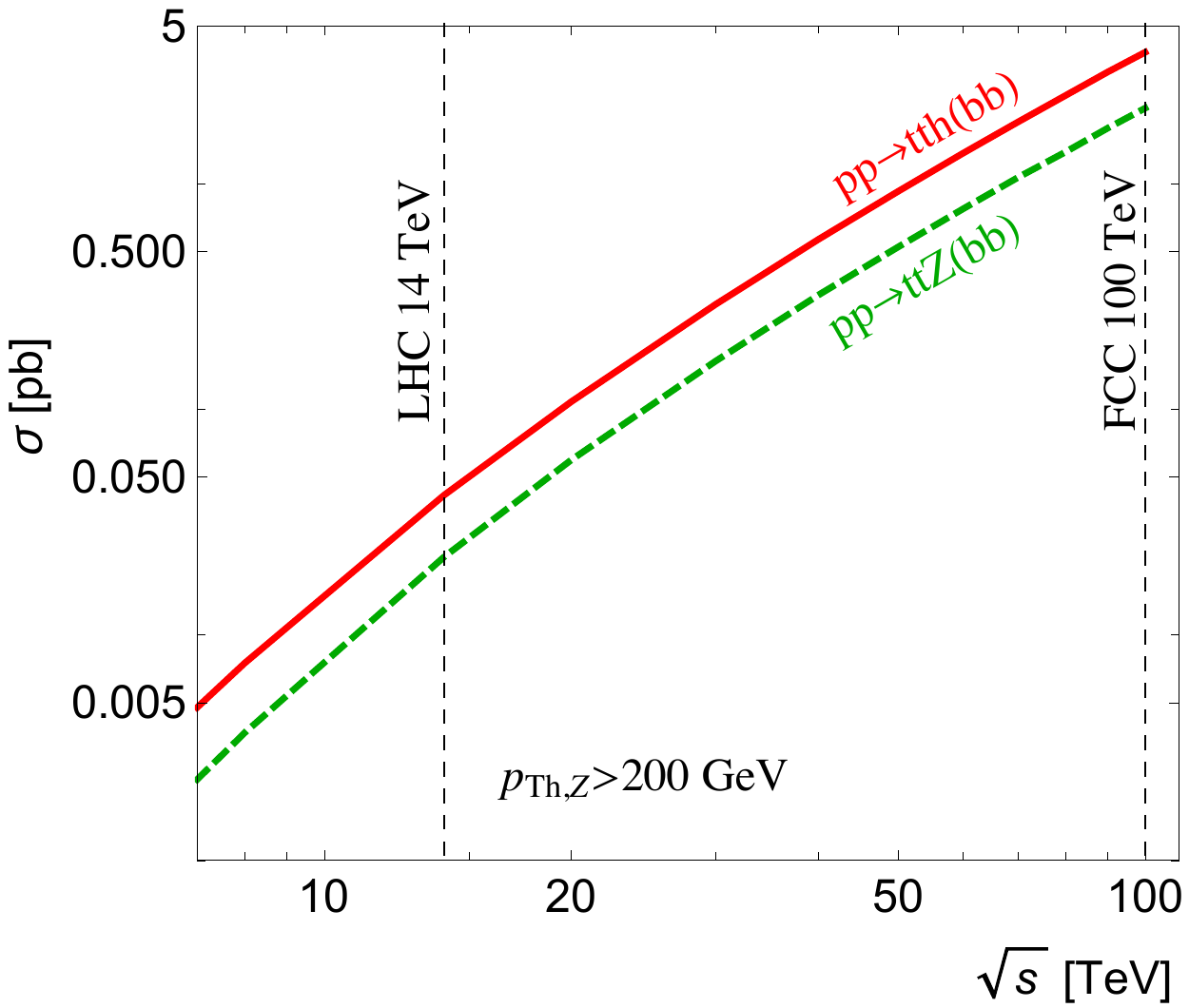}
    \caption{Cross-section for $pp\to t\bar{t}h$ and $pp\to t\bar{t}Z$ production at the parton level as a function of the $pp$ collider energy.
    We require the Higgs and Z bosons in the boosted regime, $p_{Th,Z}>200$~GeV, and account for their branching ratios to a bottom-quark pair, $\mathcal{BR}(h,Z\to b\bar{b})$. Top quarks are set stable.
    \label{fig:ttX-xsection} }
\end{figure}

The Higgs-top CP-phase measurement would render remarkable gains at a future 100~TeV collider due to the immensely increased statistics. In~\autoref{fig:ttX-xsection}, we show the cross-section for $pp\to t\bar{t}h$ and $pp\to t\bar{t}Z$ production  as a function of the collider energy. We require the Higgs and $Z$ bosons in the boosted regime, $p_{Th,Z}>200$~GeV, and account for their branching ratios to bottom quarks, $\mathcal{BR}(h,Z\to b\bar{b})$. While the $t\bar{t}(h\to b\bar{b})$ and  $t\bar{t}(Z\to b\bar{b})$ processes are phase space suppressed at the 14~TeV LHC, with  limited production cross-sections of 0.04~pb and 0.02~pb, the 100~TeV collider would result in  one hundred-fold enhancement, with a cross-section of 3.8~pb and 2.1~pb, respectively. Considering the leptonic top pair decay, this corresponds to an uplift in the number of events for the $t\bar{t}h$ signal from $5.8\times 10^3$  at the HL-LHC with 3~ab$^{-1}$ to $5.5\times 10^6$ at 100~TeV with 30~ab$^{-1}$.
This estimate shows that the 100~TeV FCC, with a combination of the increased energy and luminosity, can push further forward  precision measurements with the $t\bar{t}h$ channel. Instead of focusing on the semi-leptonic top pair mode, as in Ref.~\cite{Plehn:2015cta}, we explore the di-leptonic $t\bar{t}h$ system. In addition to the extra background suppression, this channel provides a better probe to the top polarization, using the charged leptons. The larger spin analyzing power associated with the charged leptons results in the stronger CP-violation observables, such as $\Delta\phi_{\ell\ell}^{t\bar{t}}$, strengthening our CP-sensitivity.

\begin{figure}[!t]
    \centering
    \includegraphics[width=0.5\textwidth]{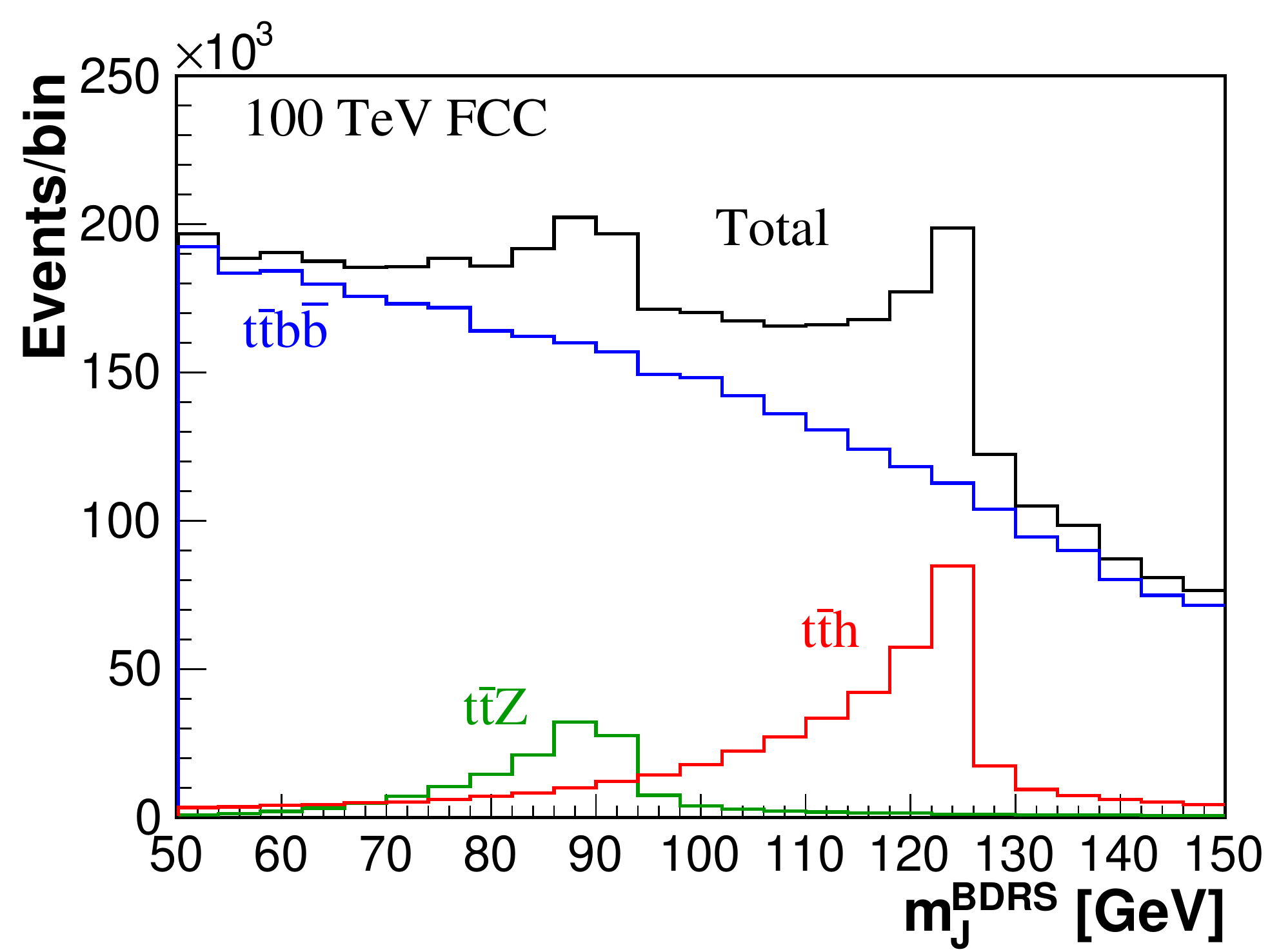}
    \caption{
    The invariant mass distributions of the signal and background for the BDRS tagged fat-jet $m_J^{\rm BDRS}$ at the 100~TeV FCC-hh.
    We show $t\bar th$  signal (red), the $t\bar{t}b\bar{b}$ (blue), and $t\bar tZ$ (green) in a non-stacked format. The full stacked result is also presented (black). We assume 30 ab$^{-1}$ of the integrated luminosity.
    \label{fig:mj100}}
\end{figure}

\begin{figure*}[t]
\begin{center}
\includegraphics[width=0.33\textwidth,clip]{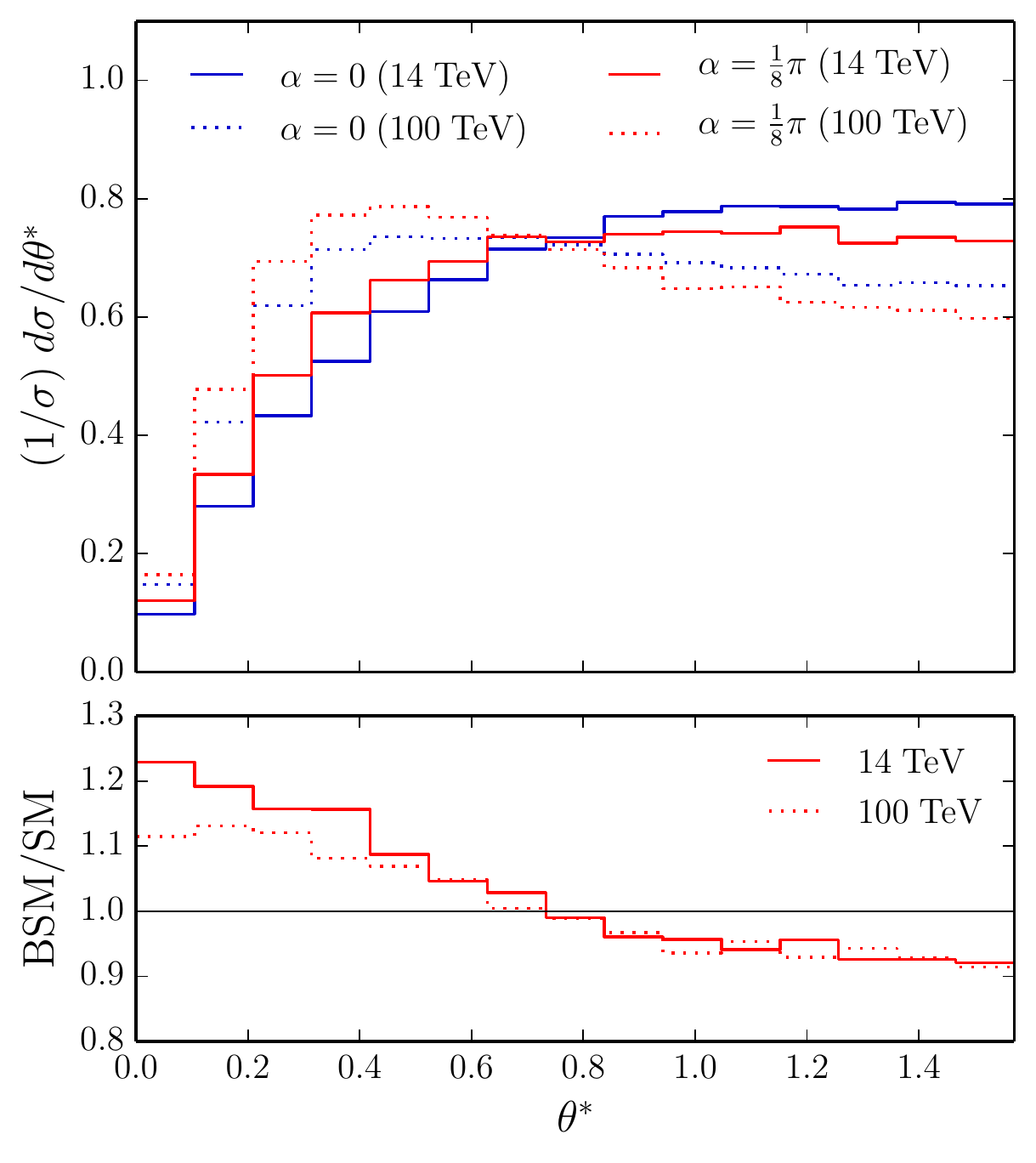}\hspace*{-0.17cm}
\includegraphics[width=0.33\textwidth,clip]{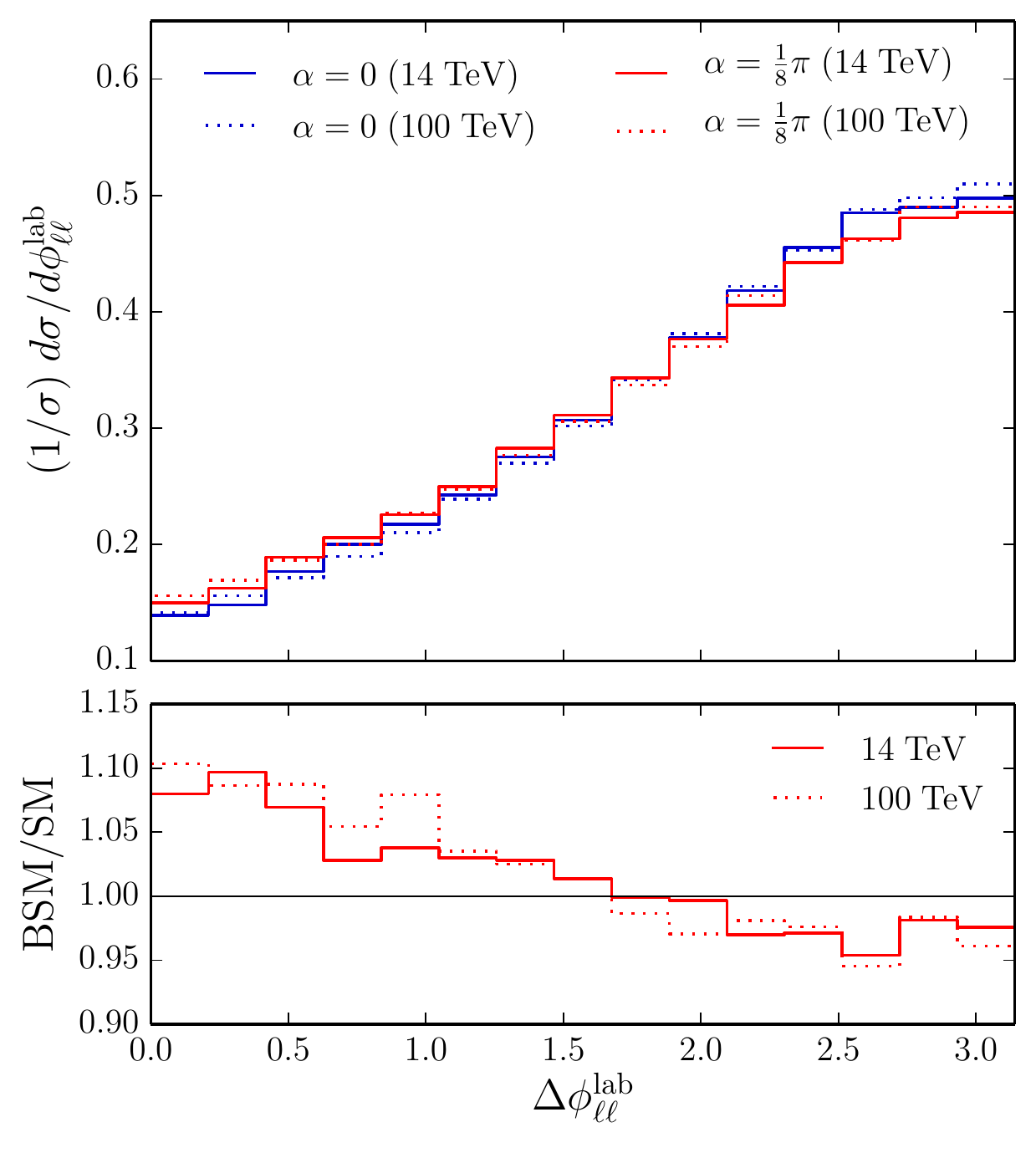}\hspace*{-0.17cm}
\includegraphics[width=0.33\textwidth,clip]{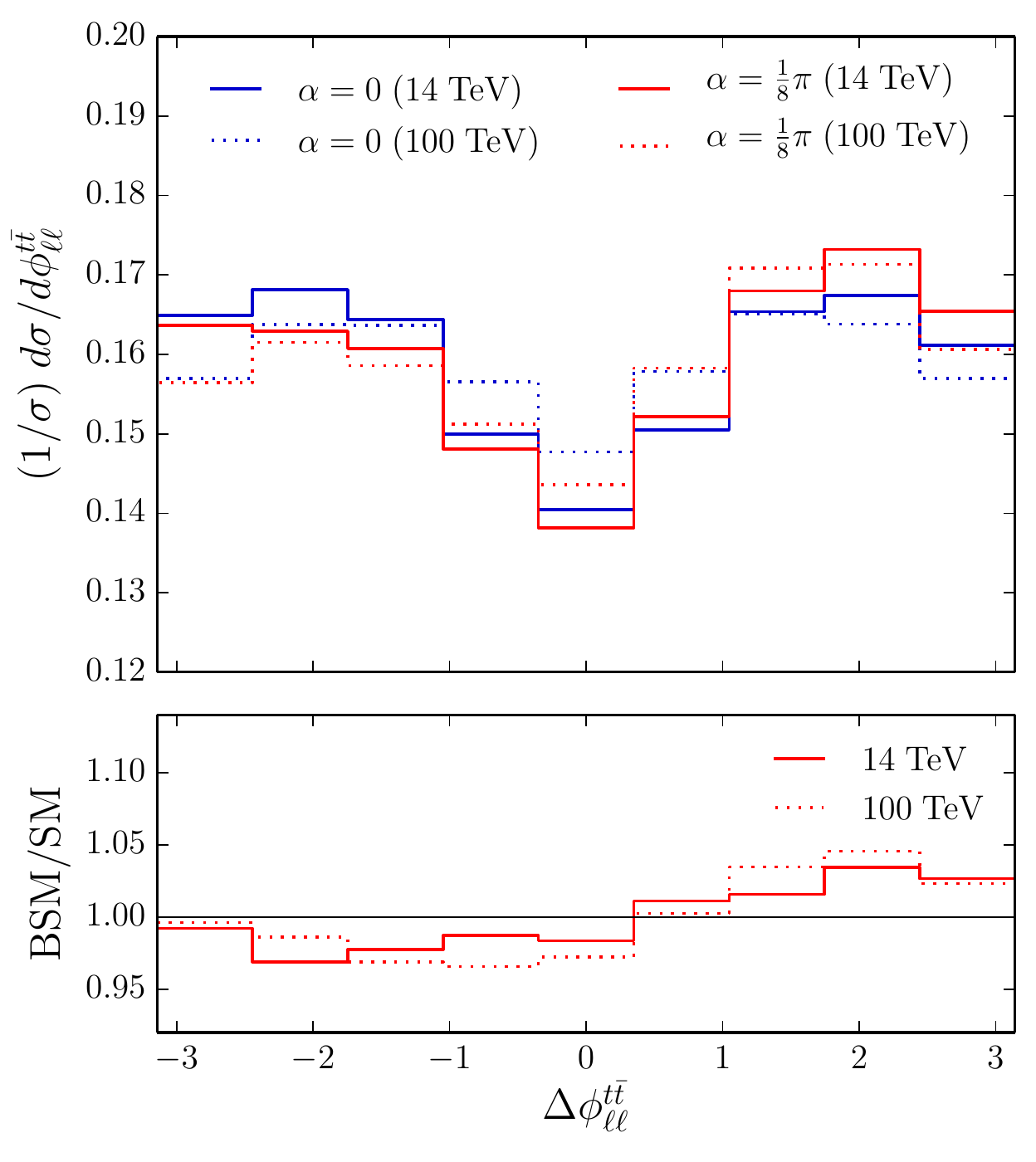}
\caption{Comparisons between 14~TeV and 100~TeV distributions of $\theta^\ast$ (left), $\Delta \phi^\text{lab}_{\ell \ell }$ (middle), and $\Delta \phi^{t \bar{t}}_{\ell \ell }$ (right) after resolving the combinatorial problems described in~\autoref{tab:Cutflow1} and~\autoref{tab:Cutflow100tev},  respectively.
}\label{ttg_FR_combo}
\end{center}
\end{figure*}

We begin our discussion with the fat-jet invariant mass $m_J^{\rm BDRS}$ distribution for the signal and background samples at the 100~TeV FCC-hh with 30 ab$^{-1}$ of data as shown~\autoref{fig:mj100} (for the full hadron level analysis). Note the $\mathcal{O}(10^3)$ fold enhancement in event rate compared to that in~\autoref{fig:mj} for the 14~TeV. The full stacked histogram is presented in black.
CP sensitive angular variables are shown in~\autoref{ttg_FR_combo}, where we present distributions of $\theta^\ast$ (left), $\Delta \phi^\text{lab}_{\ell \ell }$ (middle), and $\Delta \phi^{t \bar{t}}_{\ell \ell }$ (right) at both 14~TeV and 100~TeV for comparison.
In the laboratory frame, $\Delta \phi^\text{lab}_{\ell \ell }$ distributions look similar, while
$\theta^\ast$ and $\Delta \phi^{t \bar{t}}_{\ell \ell }$ tend to be slightly forward or backward in the $t\bar t$ rest frame.
However, the ratio of new physics contribution to the SM prediction remain similar, as shown in the bottom of each panel.

\begin{table*}[t]
\begin{center}
\setlength{\tabcolsep}{0.9mm}
\renewcommand{\arraystretch}{1.4}
\scalebox{1.0}{
\hspace*{-20pt}
\begin{tabular}{|c||c|c|c||c|c|}
\hline
    cuts                                                                                                     & $t \bar{t} h $   & $t \bar{t} b \bar{b} $           & $t \bar{t} Z $         & $\sigma$        \\  \hline \hline
~~$N_{h}= 1$, $4b$-tags, $p_T^\ell  > 20\ {\rm GeV}$, $|\eta^\ell | < 2.5$ ~~ &  \multirow{2}{*}{ ~$21.5$~ }      &   \multirow{2}{*}{ ~351~ }         &  \multirow{2}{*}{~6.93~}  &  \multirow{2}{*}{ ~61.6~ }       \\
$p_T^j  > 30\ {\rm GeV}$, $|\eta^j| < 2.5$, $N_{j} \geqslant 2$, $N_{\ell} = 2$    &                                               &                                            &                                        &                 \\  \hline
$50~\ {\rm GeV} < m_J^\text{BDRS} < 150\ {\rm GeV}$                                                                       & $17.7$                               & $177$                                 & $6.15$            &     $70.7$  \\  \hline
Resolving combinatorics                                                                       & $14.0$                               & $116$                               & $5.11$                       &    $68.4$    \\  \hline 
\end{tabular}}
\end{center}
\caption{Cumulative cut-flow table showing  cross-section in fb for  $t\bar{t}h$ signal ($\kappa_t=1,\alpha = 0$) and leading backgrounds $t\bar{t}b\bar{b}$ and $t\bar{t}Z$ at a 100~TeV future collider. The signal significances ($\sigma$) are calculated for a luminosity of 30~$\rm{ab}^{-1}$.}
\label{tab:Cutflow100tev}
\end{table*}

\begin{figure*}
	\centering
	\includegraphics[width=\textwidth]{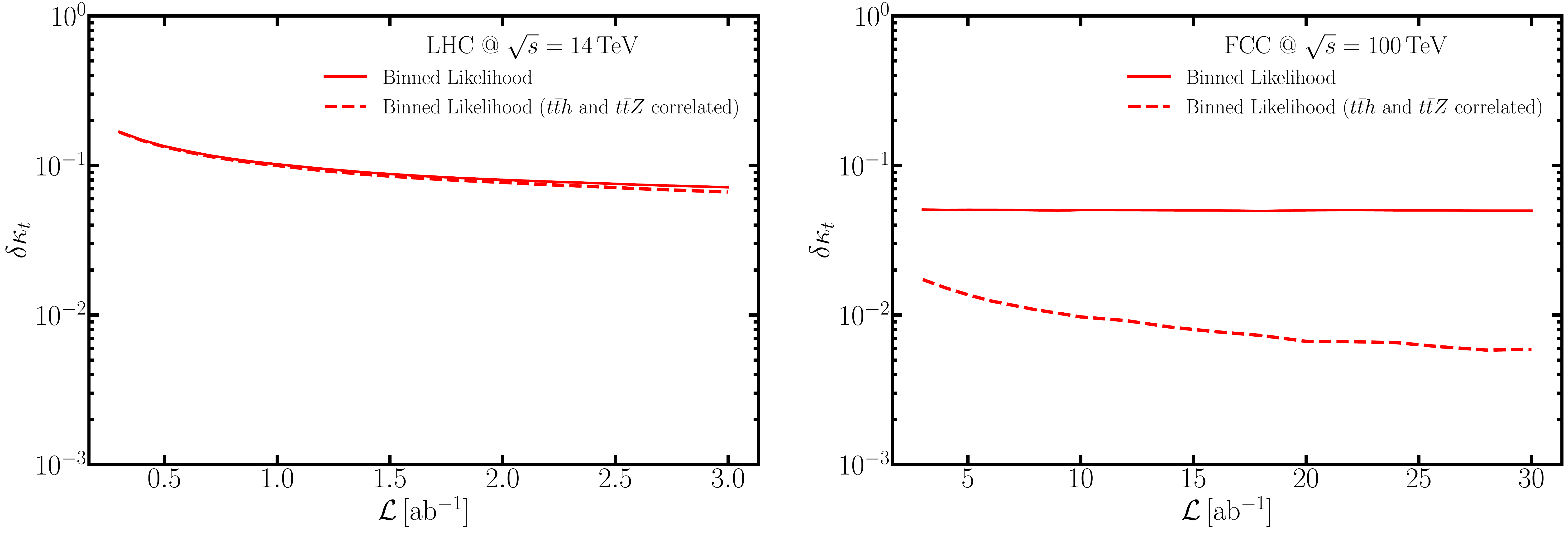}
	\caption{The precision on $\kappa_t$ assuming $\alpha = 0$ at $\sqrt{s}=14$ TeV (left) and $\sqrt{s}=100$ TeV (right), using binned log-likelihood method.
    20\% systematics for $t\bar{t}b\bar{b}$ and 10\% systematics for $t\bar{t}h$ and $t\bar{t}z$ are assumed.
    For dashed curves, the uncertainties for $t\bar{t}h$ and $t\bar{t}z$ are assumed to be correlated.
	\label{fig:dkt_lumi}}
\end{figure*}

As mentioned previously, one main difference between 14~TeV LHC and 100~TeV FCC is the significant increase in the rate of signal and backgrounds. Especially both $t\bar{t}h$ and $t\bar{t}Z$ $(i)$~result in hugely improved statistics, $(ii)$~have similar production mechanisms, and $(iii)$~probe comparable energy scales.
Hence, their uncertainties are highly correlated~\cite{Plehn:2015cta}. The theoretical uncertainties in the signal cross-section, that are in the range  7-10\% at 100~TeV collider, can be depleted to approximately 1\% in terms of a ratio measurement~\cite{Plehn:2015cta}. This reduction of the uncertainties is also depicted in~\autoref{fig:dkt_lumi} for the 14~TeV LHC (left) and the 100~TeV FCC (right), where we only consider the precision on the $\kappa_t$ measurement by fixing $\alpha = 0$. We considered two different scenarios: $(1)$ binned log-likelihood (red-solid); and $(2)$ binned log-likelihood with $t\bar th$ and $t\bar tZ$ correlated in uncertainties (red-dashed). At the 14~TeV LHC, whether we consider the correlation between the uncertainties of $t\bar th$ and $t\bar tZ$ ($i.e.$ we use the same nuisance parameter for the uncertainties of $t\bar th$ and $t\bar tZ$), does not significantly affect the results, as the uncertainties are dominated by the continuum $t\bar t b\bar b$ background. However, the situation improves dramatically at the 100~TeV FCC. The scenario $(1)$ is systematically limited around $\delta \kappa_t\lesssim 5\%$ due to the 10\% systematics on the rate of the $t\bar th$.  When considering the correlation between $t\bar th$ and $t\bar tZ$ in scenario $(2)$, the $\kappa_t$ measurement improves and can reach sub-percentage precision. Note that our results for $\kappa_t$ at a 100~TeV collider, $\delta\kappa_t\lesssim 0.5-0.7\%$, are consistent with those from Ref.~\cite{Plehn:2015cta}, that explores the semi-leptonic top pair final state.

\begin{figure*}[t]
    \centering
    \includegraphics[width=\textwidth]{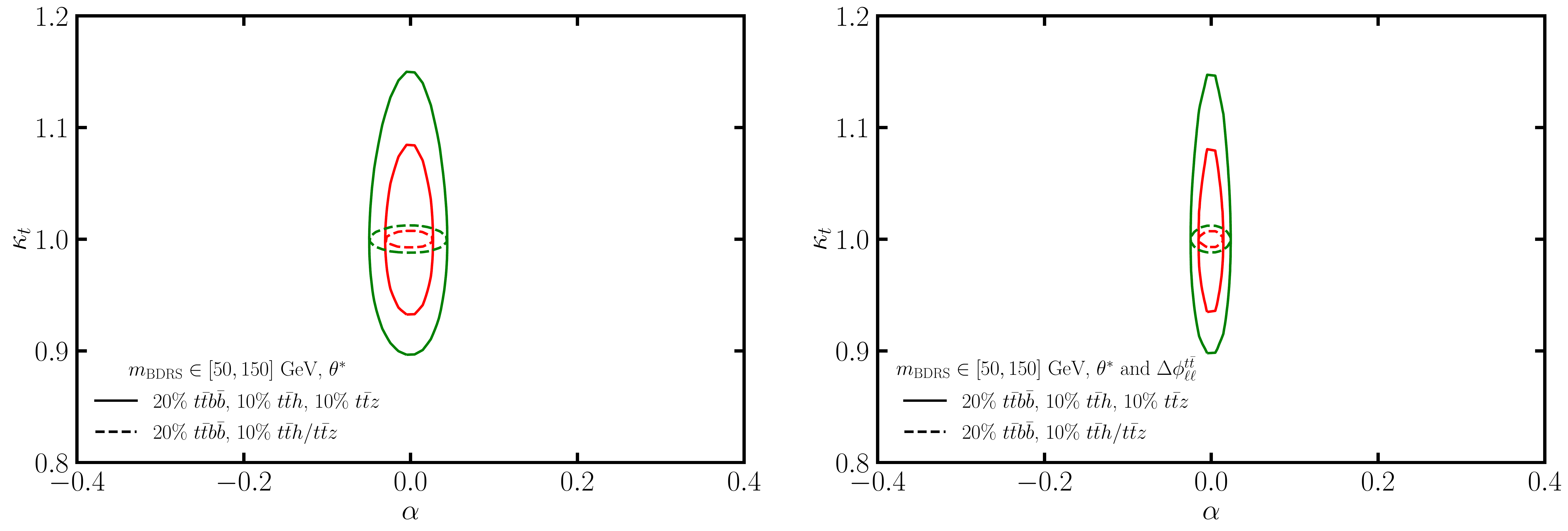}
    \caption{68\% (red) and 95\% (green) CL limits on the $\alpha$-$\kappa_t$ plane for the 100~TeV FCC with $30\ {\rm ab}^{-1}$ without (left) and with (right)  $\Delta\phi_{\ell\ell}^{t\bar t}$.
    For the solid curves, 10\% systematics is used for both $t\bar th$ and $t\bar tZ$ individually, while for the dashed curves, the uncertainties for $t\bar t h$ and $t\bar t Z$ are assumed to be correlated. 20\% systematics is used for $t\bar tb\bar b$ for both scenarios.}
    \label{fig:alpha_kt_100TeV}
\end{figure*}

In light of the aforementioned improvements on the $\kappa_t$ sensitivity, we perform a similar analysis considering both $\kappa_t$ and $\alpha$. With the uplifted cross-section and enlarged luminosity, the 100~TeV FCC can boost the sensitivities on ($\alpha$, $\kappa_t$), using the binned log-likelihood method, as summarized in~\autoref{fig:alpha_kt_100TeV}.
We choose a wide mass window, $m_{\rm BDRS} \in [50, 150]$ GeV for better control of the continuum $t\bar{t}b\bar{b}$ background, along with $\theta^\ast$ in the left panel. In both panels, the solid curves correspond to the case with 20\% systematics for $t\bar tb\bar b$ and 10\% systematics for $t\bar t h$ and $t\bar tZ$, while we assume $t\bar t h$ and $t\bar t Z$ uncertainties are correlated for the dashed curves. It is clear that, at a high luminosity,  the solid curves are limited by the systematic uncertainties, similarly to the solid red line scenario in the right panel of~\autoref{fig:dkt_lumi}.
However, by assuming that the systematics of $t\bar th$ is correlated with $t\bar tZ$, the precision can be improved, as shown by the dashed curves, which can achieve $\delta\kappa_t\lesssim  1\%$ and $| \alpha| \lesssim 3^\circ$ at 95\%~CL.

Finally, extending the analysis to the $(m_J^{\rm BDRS},\theta^*,\Delta\phi_{\ell\ell}^{t\bar t})$ plane, we find that the CP-odd observable $\Delta\phi_{\ell\ell}^{t \bar t}$ brings additional improvement on the measurement of $\alpha$ by a factor of 2, $| \alpha| \lesssim 1.5^\circ$, as shown in the right panel of~\autoref{fig:alpha_kt_100TeV}, which highlights the importance of the CP-odd observable in the $t\bar t$ rest frame.

\section{Summary}
\label{sec:summary}
The discovery of the Higgs boson at the LHC jump-started a comprehensive program of precision measurements for the Higgs couplings. In this context, the direct measurement of the Higgs-top coupling strength and CP-phase would have a significant impact on our understanding of the Yukawa sector and possible new sources of CP violation.
In this paper, we have examined the direct probe of the top quark Yukawa coupling and the Higgs-top CP-structure in the $t\bar t h$ production, with the Higgs boson decaying to a bottom pair and top-quarks in the di-leptonic mode. We have utilized several state-of-the-art strategies to reconstruct the final state with the missing transverse momentum and to control systematic uncertainties. We take advantage of the BDRS algorithm to tag the boosted Higgs, and exploit the $M_2$-assisted reconstruction to compute observables sensitive to the CP-phase at the $t\bar{t}$ rest frame. Our log-likelihood analysis, using the side-band control region,  takes full advantage of the shape information of the signal and background. We have shown that the proposed analysis significantly reduces the uncertainty in the CP-phase measurement and the Higgs-top Yukawa coupling. Our results show that the Higgs-top CP-phase ($\alpha$) can be probed up to $| \alpha | \lesssim  36^\circ$ and the top Yukawa ($\kappa_t$) up to $\sim 20\%$ accuracy (95\%~CL) at the HL-LHC,  as shown in~\autoref{fig:alpha_kt_14TeV}. A similar analysis at a 100~TeV future collider further improves the precision on the coupling modifier and CP-phase to $\delta\kappa_t\lesssim  1\%$ and $| \alpha| \lesssim 3^\circ$, respectively, as shown in~\autoref{fig:alpha_kt_100TeV}. We find that the CP-odd observable $\Delta\phi_{\ell\ell}^{t \bar t}$ augments the precision by a factor of 2, $| \alpha| \lesssim 1.5^\circ$. We note these limits represent only an upper bound, that can be further enhanced via the combination of the other relevant top-quark and Higgs decays from the $t\bar{t}h$ production.

\section*{Acknowledgements}
\label{sec:acknowledgements}
We thank Joseph Haley, Youngjoon Kwon, and Yue Xu for useful discussion on the control regions. DG and YW thank the U.S.~Department of Energy for the financial support, under grant number DE-SC 0016013. KK acknowledges support from the US DOE, Office of Science under contract DE-SC0021447.
JK is supported by the National Research Foundation of Korea (NRF) grant funded by the Korea government (MSIT) (No. 2021R1C1C1005076).

\bibliographystyle{JHEP}
\bibliography{references}

\providecommand{\href}[2]{#2}\begingroup\raggedright\begin{thebibliography}{10}

\bibitem{Sakharov:1967dj}
A.~D. Sakharov, \emph{{Violation of CP Invariance, C asymmetry, and baryon
  asymmetry of the universe}},
  \href{http://dx.doi.org/10.1070/PU1991v034n05ABEH002497}{\emph{Pisma Zh.
  Eksp. Teor. Fiz.} {\bf 5} (1967) 32--35}.

\bibitem{Buchmuller:1985jz}
W.~Buchmuller and D.~Wyler, \emph{{Effective Lagrangian Analysis of New
  Interactions and Flavor Conservation}},
  \href{http://dx.doi.org/10.1016/0550-3213(86)90262-2}{\emph{Nucl. Phys.} {\bf
  B268} (1986) 621--653}.

\bibitem{Grzadkowski:2010es}
B.~Grzadkowski, M.~Iskrzynski, M.~Misiak and J.~Rosiek, \emph{{Dimension-Six
  Terms in the Standard Model Lagrangian}},
  \href{http://dx.doi.org/10.1007/JHEP10(2010)085}{\emph{JHEP} {\bf 10} (2010)
  085}, [\href{https://arxiv.org/abs/1008.4884}{{\tt 1008.4884}}].

\bibitem{Buckley:2015vsa}
M.~R. Buckley and D.~Goncalves, \emph{{Boosting the Direct CP Measurement of
  the Higgs-Top Coupling}},
  \href{http://dx.doi.org/10.1103/PhysRevLett.116.091801}{\emph{Phys. Rev.
  Lett.} {\bf 116} (2016) 091801},
  [\href{https://arxiv.org/abs/1507.07926}{{\tt 1507.07926}}].

\bibitem{Brod:2013cka}
J.~Brod, U.~Haisch and J.~Zupan, \emph{{Constraints on CP-violating Higgs
  couplings to the third generation}},
  \href{http://dx.doi.org/10.1007/JHEP11(2013)180}{\emph{JHEP} {\bf 11} (2013)
  180}, [\href{https://arxiv.org/abs/1310.1385}{{\tt 1310.1385}}].

\bibitem{Dolan:2014upa}
M.~J. Dolan, P.~Harris, M.~Jankowiak and M.~Spannowsky, \emph{{Constraining
  $CP$-violating Higgs Sectors at the LHC using gluon fusion}},
  \href{http://dx.doi.org/10.1103/PhysRevD.90.073008}{\emph{Phys. Rev. D} {\bf
  90} (2014) 073008}, [\href{https://arxiv.org/abs/1406.3322}{{\tt
  1406.3322}}].

\bibitem{Englert:2012xt}
C.~Englert, D.~Goncalves-Netto, K.~Mawatari and T.~Plehn, \emph{{Higgs Quantum
  Numbers in Weak Boson Fusion}},
  \href{http://dx.doi.org/10.1007/JHEP01(2013)148}{\emph{JHEP} {\bf 01} (2013)
  148}, [\href{https://arxiv.org/abs/1212.0843}{{\tt 1212.0843}}].

\bibitem{Bernlochner:2018opw}
F.~U. Bernlochner, C.~Englert, C.~Hays, K.~Lohwasser, H.~Mildner, A.~Pilkington
  et~al., \emph{{Angles on CP-violation in Higgs boson interactions}},
  \href{http://dx.doi.org/10.1016/j.physletb.2019.01.043}{\emph{Phys. Lett. B}
  {\bf 790} (2019) 372--379}, [\href{https://arxiv.org/abs/1808.06577}{{\tt
  1808.06577}}].

\bibitem{Englert:2019xhk}
C.~Englert, P.~Galler, A.~Pilkington and M.~Spannowsky, \emph{{Approaching
  robust EFT limits for CP-violation in the Higgs sector}},
  \href{http://dx.doi.org/10.1103/PhysRevD.99.095007}{\emph{Phys. Rev. D} {\bf
  99} (2019) 095007}, [\href{https://arxiv.org/abs/1901.05982}{{\tt
  1901.05982}}].

\bibitem{Gritsan:2020pib}
A.~V. Gritsan, J.~Roskes, U.~Sarica, M.~Schulze, M.~Xiao and Y.~Zhou,
  \emph{{New features in the JHU generator framework: constraining Higgs boson
  properties from on-shell and off-shell production}},
  \href{http://dx.doi.org/10.1103/PhysRevD.102.056022}{\emph{Phys. Rev. D} {\bf
  102} (2020) 056022}, [\href{https://arxiv.org/abs/2002.09888}{{\tt
  2002.09888}}].

\bibitem{Bahl:2020wee}
H.~Bahl, P.~Bechtle, S.~Heinemeyer, J.~Katzy, T.~Klingl, K.~Peters et~al.,
  \emph{{Indirect $\mathcal{CP}$ probes of the Higgs-top-quark interaction:
  current LHC constraints and future opportunities}},
  \href{http://dx.doi.org/10.1007/JHEP11(2020)127}{\emph{JHEP} {\bf 11} (2020)
  127}, [\href{https://arxiv.org/abs/2007.08542}{{\tt 2007.08542}}].

\bibitem{Ellis:2013yxa}
J.~Ellis, D.~S. Hwang, K.~Sakurai and M.~Takeuchi, \emph{{Disentangling
  Higgs-Top Couplings in Associated Production}},
  \href{http://dx.doi.org/10.1007/JHEP04(2014)004}{\emph{JHEP} {\bf 04} (2014)
  004}, [\href{https://arxiv.org/abs/1312.5736}{{\tt 1312.5736}}].

\bibitem{Boudjema:2015nda}
F.~Boudjema, R.~M. Godbole, D.~Guadagnoli and K.~A. Mohan, \emph{{Lab-frame
  observables for probing the top-Higgs interaction}},
  \href{http://dx.doi.org/10.1103/PhysRevD.92.015019}{\emph{Phys. Rev. D} {\bf
  92} (2015) 015019}, [\href{https://arxiv.org/abs/1501.03157}{{\tt
  1501.03157}}].

\bibitem{Buckley:2015ctj}
M.~R. Buckley and D.~Goncalves, \emph{{Constraining the Strength and CP
  Structure of Dark Production at the LHC: the Associated Top-Pair Channel}},
  \href{http://dx.doi.org/10.1103/PhysRevD.93.034003}{\emph{Phys. Rev. D} {\bf
  93} (2016) 034003}, [\href{https://arxiv.org/abs/1511.06451}{{\tt
  1511.06451}}].

\bibitem{Gritsan:2016hjl}
A.~V. Gritsan, R.~R\"ontsch, M.~Schulze and M.~Xiao, \emph{{Constraining
  anomalous Higgs boson couplings to the heavy flavor fermions using matrix
  element techniques}},
  \href{http://dx.doi.org/10.1103/PhysRevD.94.055023}{\emph{Phys. Rev. D} {\bf
  94} (2016) 055023}, [\href{https://arxiv.org/abs/1606.03107}{{\tt
  1606.03107}}].

\bibitem{Goncalves:2016qhh}
D.~Goncalves and D.~Lopez-Val, \emph{{Pseudoscalar searches with dileptonic
  tops and jet substructure}},
  \href{http://dx.doi.org/10.1103/PhysRevD.94.095005}{\emph{Phys. Rev. D} {\bf
  94} (2016) 095005}, [\href{https://arxiv.org/abs/1607.08614}{{\tt
  1607.08614}}].

\bibitem{AmorDosSantos:2017ayi}
S.~Amor Dos~Santos et~al., \emph{{Probing the CP nature of the Higgs coupling
  in $t{\bar t}h$ events at the LHC}},
  \href{http://dx.doi.org/10.1103/PhysRevD.96.013004}{\emph{Phys. Rev. D} {\bf
  96} (2017) 013004}, [\href{https://arxiv.org/abs/1704.03565}{{\tt
  1704.03565}}].

\bibitem{Azevedo:2017qiz}
D.~Azevedo, A.~Onofre, F.~Filthaut and R.~Gon\c{c}alo, \emph{{CP tests of Higgs
  couplings in $t\bar{t}h$ semileptonic events at the LHC}},
  \href{http://dx.doi.org/10.1103/PhysRevD.98.033004}{\emph{Phys. Rev. D} {\bf
  98} (2018) 033004}, [\href{https://arxiv.org/abs/1711.05292}{{\tt
  1711.05292}}].

\bibitem{Goncalves:2018agy}
D.~Gon\c{c}alves, K.~Kong and J.~H. Kim, \emph{{Probing the top-Higgs Yukawa CP
  structure in dileptonic $ t\overline{t}h $ with M$_{2}$-assisted
  reconstruction}},
  \href{http://dx.doi.org/10.1007/JHEP06(2018)079}{\emph{JHEP} {\bf 06} (2018)
  079}, [\href{https://arxiv.org/abs/1804.05874}{{\tt 1804.05874}}].

\bibitem{ATLAS:2018mme}
{\scshape ATLAS} collaboration, M.~Aaboud et~al., \emph{{Observation of Higgs
  boson production in association with a top quark pair at the LHC with the
  ATLAS detector}},
  \href{http://dx.doi.org/10.1016/j.physletb.2018.07.035}{\emph{Phys. Lett. B}
  {\bf 784} (2018) 173--191}, [\href{https://arxiv.org/abs/1806.00425}{{\tt
  1806.00425}}].

\bibitem{CMS:2018uxb}
{\scshape CMS} collaboration, A.~M. Sirunyan et~al., \emph{{Observation of
  $\mathrm{t\overline{t}}$H production}},
  \href{http://dx.doi.org/10.1103/PhysRevLett.120.231801}{\emph{Phys. Rev.
  Lett.} {\bf 120} (2018) 231801},
  [\href{https://arxiv.org/abs/1804.02610}{{\tt 1804.02610}}].

\bibitem{Bortolato:2020zcg}
B.~Bortolato, J.~F. Kamenik, N.~Ko\v{s}nik and A.~Smolkovi\v{c},
  \emph{{Optimized probes of $CP$ -odd effects in the $t \bar t h$ process at
  hadron colliders}},
  \href{http://dx.doi.org/10.1016/j.nuclphysb.2021.115328}{\emph{Nucl. Phys. B}
  {\bf 964} (2021) 115328}, [\href{https://arxiv.org/abs/2006.13110}{{\tt
  2006.13110}}].

\bibitem{MammenAbraham:2021ssc}
R.~Mammen~Abraham, D.~Gon\c{c}alves, T.~Han, S.~C.~I. Leung and H.~Qin,
  \emph{{Directly Probing the Higgs-top Coupling at High Scales}},
  \href{https://arxiv.org/abs/2106.00018}{{\tt 2106.00018}}.

\bibitem{ATLAS:2020ior}
{\scshape ATLAS} collaboration, G.~Aad et~al., \emph{{$CP$ Properties of Higgs
  Boson Interactions with Top Quarks in the $t\bar{t}H$ and $tH$ Processes
  Using $H \rightarrow \gamma\gamma$ with the ATLAS Detector}},
  \href{http://dx.doi.org/10.1103/PhysRevLett.125.061802}{\emph{Phys. Rev.
  Lett.} {\bf 125} (2020) 061802},
  [\href{https://arxiv.org/abs/2004.04545}{{\tt 2004.04545}}].

\bibitem{CMS:2020cga}
{\scshape CMS} collaboration, A.~M. Sirunyan et~al., \emph{{Measurements of
  $\mathrm{t\bar{t}}H$ Production and the CP Structure of the Yukawa
  Interaction between the Higgs Boson and Top Quark in the Diphoton Decay
  Channel}},
  \href{http://dx.doi.org/10.1103/PhysRevLett.125.061801}{\emph{Phys. Rev.
  Lett.} {\bf 125} (2020) 061801},
  [\href{https://arxiv.org/abs/2003.10866}{{\tt 2003.10866}}].

\bibitem{Cepeda:2019klc}
M.~Cepeda et~al., \emph{{Report from Working Group 2}: {Higgs Physics at the
  HL-LHC and HE-LHC}},
  \href{http://dx.doi.org/10.23731/CYRM-2019-007.221}{\emph{CERN Yellow Rep.
  Monogr.} {\bf 7} (2019) 221--584},
  [\href{https://arxiv.org/abs/1902.00134}{{\tt 1902.00134}}].

\bibitem{ATLAS:2017fak}
{\scshape ATLAS} collaboration, M.~Aaboud et~al., \emph{{Search for the
  standard model Higgs boson produced in association with top quarks and
  decaying into a $b\bar{b}$ pair in $pp$ collisions at $\sqrt{s}$ = 13 TeV
  with the ATLAS detector}},
  \href{http://dx.doi.org/10.1103/PhysRevD.97.072016}{\emph{Phys. Rev. D} {\bf
  97} (2018) 072016}, [\href{https://arxiv.org/abs/1712.08895}{{\tt
  1712.08895}}].

\bibitem{CMS:2018hnq}
{\scshape CMS} collaboration, A.~M. Sirunyan et~al., \emph{{Search for $
  \mathrm{t}\overline{\mathrm{t}}\mathrm{H} $ production in the $ \mathrm{H}\to
  \mathrm{b}\overline{\mathrm{b}} $ decay channel with leptonic $
  \mathrm{t}\overline{\mathrm{t}} $ decays in proton-proton collisions at $
  \sqrt{s}=13 $ TeV}},
  \href{http://dx.doi.org/10.1007/JHEP03(2019)026}{\emph{JHEP} {\bf 03} (2019)
  026}, [\href{https://arxiv.org/abs/1804.03682}{{\tt 1804.03682}}].

\bibitem{Plehn:2015cta}
M.~L. Mangano, T.~Plehn, P.~Reimitz, T.~Schell and H.-S. Shao, \emph{{Measuring
  the Top Yukawa Coupling at 100 TeV}},
  \href{http://dx.doi.org/10.1088/0954-3899/43/3/035001}{\emph{J. Phys. G} {\bf
  43} (2016) 035001}, [\href{https://arxiv.org/abs/1507.08169}{{\tt
  1507.08169}}].

\bibitem{Bernreuther:2010ny}
W.~Bernreuther and Z.-G. Si, \emph{{Distributions and correlations for top
  quark pair production and decay at the Tevatron and LHC.}},
  \href{http://dx.doi.org/10.1016/j.nuclphysb.2010.05.001}{\emph{Nucl. Phys. B}
  {\bf 837} (2010) 90--121}, [\href{https://arxiv.org/abs/1003.3926}{{\tt
  1003.3926}}].

\bibitem{Burns:2008va}
M.~Burns, K.~Kong, K.~T. Matchev and M.~Park, \emph{{Using Subsystem MT2 for
  Complete Mass Determinations in Decay Chains with Missing Energy at Hadron
  Colliders}},
  \href{http://dx.doi.org/10.1088/1126-6708/2009/03/143}{\emph{JHEP} {\bf 03}
  (2009) 143}, [\href{https://arxiv.org/abs/0810.5576}{{\tt 0810.5576}}].

\bibitem{Barr:2011xt}
A.~J. Barr, T.~J. Khoo, P.~Konar, K.~Kong, C.~G. Lester, K.~T. Matchev et~al.,
  \emph{{Guide to transverse projections and mass-constraining variables}},
  \href{http://dx.doi.org/10.1103/PhysRevD.84.095031}{\emph{Phys. Rev. D} {\bf
  84} (2011) 095031}, [\href{https://arxiv.org/abs/1105.2977}{{\tt
  1105.2977}}].

\bibitem{Debnath:2017ktz}
D.~Debnath, D.~Kim, J.~H. Kim, K.~Kong and K.~T. Matchev, \emph{{Resolving
  Combinatorial Ambiguities in Dilepton $t\bar t$ Event Topologies with
  Constrained $M_2$ Variables}},
  \href{http://dx.doi.org/10.1103/PhysRevD.96.076005}{\emph{Phys. Rev. D} {\bf
  96} (2017) 076005}, [\href{https://arxiv.org/abs/1706.04995}{{\tt
  1706.04995}}].

\bibitem{Kim:2017awi}
D.~Kim, K.~T. Matchev, F.~Moortgat and L.~Pape, \emph{{Testing Invisible
  Momentum Ansatze in Missing Energy Events at the LHC}},
  \href{http://dx.doi.org/10.1007/JHEP08(2017)102}{\emph{JHEP} {\bf 08} (2017)
  102}, [\href{https://arxiv.org/abs/1703.06887}{{\tt 1703.06887}}].

\bibitem{Lester:1999tx}
C.~G. Lester and D.~J. Summers, \emph{{Measuring masses of semiinvisibly
  decaying particles pair produced at hadron colliders}},
  \href{http://dx.doi.org/10.1016/S0370-2693(99)00945-4}{\emph{Phys. Lett. B}
  {\bf 463} (1999) 99--103}, [\href{https://arxiv.org/abs/hep-ph/9906349}{{\tt
  hep-ph/9906349}}].

\bibitem{Ross:2007rm}
G.~G. Ross and M.~Serna, \emph{{Mass determination of new states at hadron
  colliders}},
  \href{http://dx.doi.org/10.1016/j.physletb.2008.06.003}{\emph{Phys. Lett. B}
  {\bf 665} (2008) 212--218}, [\href{https://arxiv.org/abs/0712.0943}{{\tt
  0712.0943}}].

\bibitem{Cho:2014naa}
W.~S. Cho, J.~S. Gainer, D.~Kim, K.~T. Matchev, F.~Moortgat, L.~Pape et~al.,
  \emph{{On-shell constrained $M_2$ variables with applications to mass
  measurements and topology disambiguation}},
  \href{http://dx.doi.org/10.1007/JHEP08(2014)070}{\emph{JHEP} {\bf 08} (2014)
  070}, [\href{https://arxiv.org/abs/1401.1449}{{\tt 1401.1449}}].

\bibitem{Baringer:2011nh}
P.~Baringer, K.~Kong, M.~McCaskey and D.~Noonan, \emph{{Revisiting
  Combinatorial Ambiguities at Hadron Colliders with $M_{T2}$}},
  \href{http://dx.doi.org/10.1007/JHEP10(2011)101}{\emph{JHEP} {\bf 10} (2011)
  101}, [\href{https://arxiv.org/abs/1109.1563}{{\tt 1109.1563}}].

\bibitem{Cho:2015laa}
W.~S. Cho, J.~S. Gainer, D.~Kim, S.~H. Lim, K.~T. Matchev, F.~Moortgat et~al.,
  \emph{{OPTIMASS: A Package for the Minimization of Kinematic Mass Functions
  with Constraints}},
  \href{http://dx.doi.org/10.1007/JHEP01(2016)026}{\emph{JHEP} {\bf 01} (2016)
  026}, [\href{https://arxiv.org/abs/1508.00589}{{\tt 1508.00589}}].

\bibitem{Abada:2019lih}
{\scshape FCC} collaboration, A.~Abada et~al., \emph{{FCC Physics
  Opportunities}: {Future Circular Collider Conceptual Design Report Volume
  1}}, \href{http://dx.doi.org/10.1140/epjc/s10052-019-6904-3}{\emph{Eur. Phys.
  J. C} {\bf 79} (2019) 474}.

\bibitem{Alwall:2014hca}
J.~Alwall, R.~Frederix, S.~Frixione, V.~Hirschi, F.~Maltoni, O.~Mattelaer
  et~al., \emph{{The automated computation of tree-level and next-to-leading
  order differential cross sections, and their matching to parton shower
  simulations}}, \href{http://dx.doi.org/10.1007/JHEP07(2014)079}{\emph{JHEP}
  {\bf 07} (2014) 079}, [\href{https://arxiv.org/abs/1405.0301}{{\tt
  1405.0301}}].

\bibitem{Sjostrand:2006za}
T.~Sjostrand, S.~Mrenna and P.~Z. Skands, \emph{{PYTHIA 6.4 Physics and
  Manual}}, \href{http://dx.doi.org/10.1088/1126-6708/2006/05/026}{\emph{JHEP}
  {\bf 05} (2006) 026}, [\href{https://arxiv.org/abs/hep-ph/0603175}{{\tt
  hep-ph/0603175}}].

\bibitem{Artoisenet:2012st}
P.~Artoisenet, R.~Frederix, O.~Mattelaer and R.~Rietkerk, \emph{{Automatic
  spin-entangled decays of heavy resonances in Monte Carlo simulations}},
  \href{http://dx.doi.org/10.1007/JHEP03(2013)015}{\emph{JHEP} {\bf 03} (2013)
  015}, [\href{https://arxiv.org/abs/1212.3460}{{\tt 1212.3460}}].

\bibitem{Butterworth:2008iy}
J.~M. Butterworth, A.~R. Davison, M.~Rubin and G.~P. Salam, \emph{{Jet
  substructure as a new Higgs search channel at the LHC}},
  \href{http://dx.doi.org/10.1103/PhysRevLett.100.242001}{\emph{Phys. Rev.
  Lett.} {\bf 100} (2008) 242001}, [\href{https://arxiv.org/abs/0802.2470}{{\tt
  0802.2470}}].

\bibitem{Plehn:2009rk}
T.~Plehn, G.~P. Salam and M.~Spannowsky, \emph{{Fat Jets for a Light Higgs}},
  \href{http://dx.doi.org/10.1103/PhysRevLett.104.111801}{\emph{Phys. Rev.
  Lett.} {\bf 104} (2010) 111801}, [\href{https://arxiv.org/abs/0910.5472}{{\tt
  0910.5472}}].

\bibitem{Cacciari:2011ma}
M.~Cacciari, G.~P. Salam and G.~Soyez, \emph{{FastJet User Manual}},
  \href{http://dx.doi.org/10.1140/epjc/s10052-012-1896-2}{\emph{Eur. Phys. J.
  C} {\bf 72} (2012) 1896}, [\href{https://arxiv.org/abs/1111.6097}{{\tt
  1111.6097}}].

\bibitem{CERN-LHCC-2017-021}
{\scshape ATLAS} collaboration, \emph{{Technical Design Report for the ATLAS
  Inner Tracker Pixel Detector}},  Tech. Rep. CERN-LHCC-2017-021.
  ATLAS-TDR-030, CERN, Geneva, Sep, 2017.

\bibitem{ATLAS:2012am}
{\scshape ATLAS} collaboration, G.~Aad et~al., \emph{{Jet mass and substructure
  of inclusive jets in $\sqrt{s}=7$ TeV $pp$ collisions with the ATLAS
  experiment}}, \href{http://dx.doi.org/10.1007/JHEP05(2012)128}{\emph{JHEP}
  {\bf 05} (2012) 128}, [\href{https://arxiv.org/abs/1203.4606}{{\tt
  1203.4606}}].

\bibitem{Demartin:2015uha}
F.~Demartin, F.~Maltoni, K.~Mawatari and M.~Zaro, \emph{{Higgs production in
  association with a single top quark at the LHC}},
  \href{http://dx.doi.org/10.1140/epjc/s10052-015-3475-9}{\emph{Eur. Phys. J.
  C} {\bf 75} (2015) 267}, [\href{https://arxiv.org/abs/1504.00611}{{\tt
  1504.00611}}].

\bibitem{CMS:2012qbp}
{\scshape CMS} collaboration, S.~Chatrchyan et~al., \emph{{Observation of a New
  Boson at a Mass of 125 GeV with the CMS Experiment at the LHC}},
  \href{http://dx.doi.org/10.1016/j.physletb.2012.08.021}{\emph{Phys. Lett. B}
  {\bf 716} (2012) 30--61}, [\href{https://arxiv.org/abs/1207.7235}{{\tt
  1207.7235}}].

\bibitem{ATLAS:2020fcp}
{\scshape ATLAS} collaboration, G.~Aad et~al., \emph{{Measurements of $WH$ and
  $ZH$ production in the $H \rightarrow b\bar{b}$ decay channel in $pp$
  collisions at 13 TeV with the ATLAS detector}},
  \href{http://dx.doi.org/10.1140/epjc/s10052-020-08677-2}{\emph{Eur. Phys. J.
  C} {\bf 81} (2021) 178}, [\href{https://arxiv.org/abs/2007.02873}{{\tt
  2007.02873}}].

\end{thebibliography}\endgroup

\end{document}